\documentstyle[12pt,epsf]{article}
\def\be{\begin{equation}}
\def\ee{\end{equation}}
\def\bear{\begin{eqnarray}}
\def\eear{\end{eqnarray}}
\def\simlt{\stackrel{<}{{}_\sim}}
\def\simgt{\stackrel{>}{{}_\sim}}

\newcommand{\postscript}[2]{\setlength{\epsfxsize}{#2\hsize}
   \centerline{\epsfbox{#1}}}

\def\NPB#1#2#3{{\it Nucl.~Phys.} {\bf{B#1}} (19#2) #3}
\def\PLB#1#2#3{{\it Phys.~Lett.} {\bf{B#1}} (19#2) #3}
\def\PRD#1#2#3{{\it Phys.~Rev.} {\bf{D#1}} (19#2) #3}
\def\PRL#1#2#3{{\it Phys.~Rev.~Lett.} {\bf{#1}} (19#2) #3}
\def\ZPC#1#2#3{{\it Z.~Phys.} {\bf C#1} (19#2) #3}
\def\PTP#1#2#3{{\it Prog.~Theor.~Phys.} {\bf#1}  (19#2) #3}

\def\PR#1#2#3{{\it Phys.~Rep.} {\bf#1} (19#2) #3}

\def\EPJ#1#2#3{{\it Eur.~Phys.~J.} {\bf C#1} (19#2) #3}

\def\roughly#1{\raise.3ex\hbox{$#1$\kern-.75em\lower1ex\hbox{$\sim$}}}
\newcommand{\dr}{\mbox{\footnotesize$\overline{\rm DR}$~}}
\newcommand{\ms}{\mbox{\footnotesize$\overline{\rm MS}$~}}
\newcommand{\olf}{16\pi^2}
\newcommand{\tAt}{X_t}
\newcommand{\xt}{X_t}
\newcommand{\Ls}{\ln\frac{M_S^2}{Q^2}}
\newcommand{\Lt}{\ln\frac{m_t^2}{Q^2}}
\newcommand{\Lst}{\ln\frac{M_S^2}{m_t^2}}
\newcommand{\LLs}{\ln^2\frac{M_S^2}{Q^2}}
\newcommand{\LLt}{\ln^2\frac{m_t^2}{Q^2}}
\newcommand{\LLst}{\ln^2\frac{M_S^2}{m_t^2}}
\newcommand{\hAt}{{\hat{A}_t}}
\newcommand{\hmu}{{\hat{\mu}}}
\newcommand{\hxt}{{\hat{X}}_t}

\newcommand{\hyt}{{\hat{Y}}_t}

\newcommand{\bl}{\beta_\lambda}
\newcommand{\tilt}{\tilde{t}}
\newcommand{\tilb}{\tilde{b}}
\newcommand{\tilf}{\tilde{f}}

\newcommand{\sa}{s_\alpha}
\newcommand{\ca}{c_\alpha}
\newcommand{\sbe}{s_\beta}
\newcommand{\cbe}{c_\beta}
\newcommand{\epo}{\epsilon_1}
\newcommand{\ept}{\epsilon_2}
\newcommand{\lra}{\leftrightarrow}
\newcommand{\mt}{{\cal Q}_t}
\newcommand{\lt}{\ln{m_t^2\over Q^2}}
\newcommand{\ls}{\ln{M^2_S\over Q^2}}
\newcommand{\lst}{\ln{M^2_S\over m_t^2}}
\newcommand{\mst}{{\cal Q}_{\tilde{t}}}
\newcommand{\mtilt}{m_{\tilde{t}}}
\newcommand{\MQ}{M_{\widetilde{Q}}}
\newcommand{\MU}{M_{\widetilde{U}}}
\newcommand{\MD}{M_{\widetilde{D}}}
\newcommand{\nn}{\nonumber}

\newcommand{\lntq}{\ln{m_t^2\over Q^2}}

\newcommand{\gsim}{\lower.7ex\hbox{$\;\stackrel{\textstyle>}{\sim}\;$}}
\newcommand{\lsim}{\lower.7ex\hbox{$\;\stackrel{\textstyle<}{\sim}\;$}}

\def\bigint{{\displaystyle\int}}

\relax
\begin{document}
\begin{titlepage}
\begin{flushright}
IFT-UAM/CSIC-00-09\\
MADPH-00-1158\\
hep-ph/0003246 \\
\end{flushright}
\vskip 0.1in
\begin{center}{\Large\bf Complete Two-loop Dominant Corrections to the\\[-1mm]
Mass of the Lightest ${\cal CP}$-even Higgs Boson in\\[1mm]
the Minimal Supersymmetric Standard Model}
\vskip 0.3in
{Jose Ram\'on Espinosa$^a$, Ren-Jie Zhang$^b$}\\[2mm]
{\it $^a$Instituto de Matem\'aticas y F\'{\i}sica Fundamental (CSIC)\\
Serrano 113 bis, 28006 Madrid, SPAIN\\
$^b$Department of Physics, University of Wisconsin\\
1150 University Avenue, Madison Wisconsin 53706, USA}\\

{\tt espinosa@makoki.iem.csic.es,~rjzhang@pheno.physics.wisc.edu}\\

\end{center}
\vskip.5cm
\begin{center}
{\bf Abstract}
\end{center}
\begin{quote}
Using an effective potential approach,
we compute two-loop radiative corrections to the
MSSM lightest ${\cal CP}$-even 
Higgs boson mass $M_{h^0}$ to ${\cal O}(\alpha_t^2)$
for arbitrary left-right top-squark mixing 
and $\tan\beta$. 
We find that these corrections can increase $M_{h^0}$ by as much as
5~GeV; assuming a SUSY scale of 1~TeV, the upper bound on the Higgs boson
mass is $M_{h^0}\approx 129\pm 5$ GeV 
for the top quark pole mass $175\pm 5$ GeV. 
We also derive an analytical 
approximation formula for $M_{h^0}$ which is good 
to a precision of $\lsim 0.5$ GeV
for most of the parameter space and suitable to be further improved 
by including renormalization group 
resummation of leading and next-to-leading order
logarithmic terms. Our final compact formula admits a clear
physical interpretation: 
radiative corrections up to the two-loop level can be well approximated
by a one-loop expression with parameters evaluated at the
appropriate  scales, plus a smaller finite two-loop threshold correction term.
\end{quote}
\vfil

\begin{center}
{\it Published in Nuclear Physics {\bf B586} (2000) 3-38.}
\end{center}

\vfill
\begin{flushleft}
March, 2000 \\
Revised, April, 2001 \\
\end{flushleft}

\end{titlepage}
\setcounter{footnote}{0}
\setcounter{page}{2}
\newpage
%
\noindent

\section{Introduction}

It is difficult to overestimate the importance of the experimental
discovery of the Higgs boson. It would not only help us to elucidate 
the dynamics responsible for electroweak symmetry breaking but it will
most probably offer also an important clue as to the nature of the Physics
beyond the Standard Model (SM). 

The paradigm for this new Physics, granted that a fundamental scalar 
drives electroweak symmetry breaking, is the Minimal Supersymmetric
Standard Model (MSSM) \cite{MSSM}: the most economical extension
of the Standard Model that incorporates (softly-broken) Supersymmetry (SUSY). 
In spite of the uncertainties related to the origin of supersymmetry breaking
(and therefore of the masses of the so far undetected supersymmetric
particles),
it is well known that the MSSM predicts the existence of a light Higgs
particle with mass below about $135$ GeV (this bound depends sensitively 
on the top quark mass one uses; our present calculation intends to set a
precise and firm bound).
Unlike the case of the Standard Model (in which the mass of the Higgs
boson is an unknown parameter), the mass of the light ${\cal CP}$-even
Higgs
boson of the MSSM is calculable as a function of other masses of the
model. A precise calculation of that mass is of prime importance for Higgs
searches at LEP, Tevatron and the LHC, and is the topic of this paper. 

We recall at this
point that the Higgs sector of the MSSM consists of two $SU(2)$ doublets,
$H_1$ (which gives mass to down-type quarks and charged leptons) and
$H_2$  (which gives mass to up-type quarks). The vacuum expectation values
($v_{1,2}$) of these doublets break the electroweak symmetry, after which,
the Higgs spectrum contains two ${\cal CP}$-even scalars ($h^0$ and $H^0$;
with
$m_{h^0}\leq m_{H^0}$), one ${\cal CP}$-odd
pseudoscalar ($A^0$) and a pair of charged Higgses ($H^\pm$). 
At tree-level, the masses, couplings and mixing angles of these
particles are determined by one unknown
mass parameter (say $m_{A^0}$) and the parameter $\beta$, which measures
the ratio $v_2/v_1(\equiv\tan\beta)$. In the limit $m_{A^0}\gg M_Z$ all
the
Higgs particles except $h^0$ have masses $\sim m_{A^0}$ and rearrange in a
complete $SU(2)$ doublet almost decoupled from electroweak symmetry
breaking, while $h^0$ remains light with $m_{h^0}^2\leq M_Z^2
\cos^22\beta$  and has SM
properties. This bound (which applies for any value of $m_{A^0}$ and is
saturated for $m_{A^0}\gg M_Z$)
is extremely important: it represents the limit that experimental bounds
should reach to falsify the MSSM. In fact, the present experimental bound
\cite{LEPlimit} from  LEP, including the latest data with up to
$\sqrt{s}=202$ GeV, is $m_{h^0}\simgt
107.7$ GeV (for large $m_{A^0}$, case in which the SM limit is applicable; the
limit falls to $\sim 91$ GeV for smaller $m_{A^0}$), which is well above this
bound. This
is not yet conclusive evidence against the MSSM because it does not take
into account the radiatively corrected form of the mass bound. 

Radiative corrections to $m_{h^0}^2$ have been computed using three
different techniques (or combinations of them): effective potential method
\cite{radEP,CEQR,H2,RenJie,ez}, direct diagrammatic calculation
\cite{radiag,H2W,PBMZ} and
effective theory (or renormalization
group) approach \cite{CEQR,ez,radRG,CEQW,H3}. The full one-loop
radiative corrections to $m_{h^0}$ have been computed diagrammatically.  
The most important of these corrections come from
top quark/squark loops and are given by
\begin{equation}
\label{leading}
\Delta m_{h^0}^2\ =\ {3 m_t^4\over2\pi^2v^2}\ln{m_{\tilde{t}}^2\over
m_t^2}\ ,
\end{equation}
where $m_t$ is the top quark mass, $m_{\tilde{t}}$ an average top-squark
mass and $v^2\equiv v_1^2+v_2^2=(246\ {\rm GeV})^2$. This correction can
be very
large if $m_{\tilde{t}}\gg m_t$, and in such case $m_{h^0}$ can evade
easily the current
experimental lower bound. This important ${\cal
O}(\alpha_t)$ logarithmic correction to the dimensionless ratio
$m_{h^0}^2/m_t^2$ [here $\alpha_t\equiv h_t^2/(4 \pi)$, 
where $h_t$ is the top-quark Yukawa coupling] can be most easily
reproduced using renormalization group (RG) techniques. In addition, 
there is a finite
(non-logarithmic) correction which may also be important, and which
depends on the details of the top-squark spectrum. This correction is 
(assuming again for simplicity degenerate soft masses for the top-squarks)
\be
\label{threshold1}
\Delta m_{h^0}^2\ =\ {3m_t^4\over2\pi^2 v^2}
\left({\tAt^2\over\mtilt^2}
-{\tAt^4\over12\mtilt^4}\right)\ ,
\ee
where $\tAt=A_t+\mu\cot\beta$ is the top-squark mixing parameter, 
$A_t$ the soft trilinear coupling associated to the top-Yukawa term in
the superpotential and $\mu$ the supersymmetric Higgs mass parameter.
Correction (\ref{threshold1}) is maximized for  $\tAt^2=6 m_{\tilde{t}}^2$
(the so-called `maximal-mixing' case).
When using one-loop equations like (\ref{leading}) and (\ref{threshold1})
to compute the Higgs mass one has to decide whether to use on-shell (OS) 
or running values for the mass parameters  that enter such
formulae (and if the latter, at which scale to evaluate them). The
difference between two such choices is of higher order, but can be
non-negligible, especially because of the $m_t^4$-dependence
of $\Delta m_{h^0}^2$. Although RG
techniques can be used to make an educated guess of the scale at which 
those mass parameters should be evaluated (see {\it e.g.} \cite{CEQW}), 
a precise
answer to such questions could only be unambiguously given by a two-loop
calculation like the one we perform in this paper. 

At two loops, radiative corrections to $m_{h^0}^2$ depend not only on the
large top-Yukawa coupling but also on the QCD coupling $g_3$. It is
reasonable to expect
that the dominant two-loop corrections will be of order ${\cal
O}(\alpha_s\alpha_t[\ln(m_{\tilde{t}}^2/m_t^2)]^k)$ and ${\cal
O}(\alpha_t^2[\ln(m_{\tilde{t}}^2/m_t^2)]^k), k=0,1,2
$. Terms with $k=2$ are the
two-loop leading logarithmic contributions and can be obtained by RG
techniques using one-loop RG equations; no true two-loop
calculation is required and RG resummation will take into account such
leading-logarithmic (LL) corrections to all loops. The $k=1$ terms are the
two-loop
next-to-leading-logarithmic (NTLL) corrections, 
which can be obtained (and resummed to all
loops) with two-loop RG equations. Finally, the two-loop non-logarithmic
terms ($k=0$) can be interpreted in the effective theory
language as threshold corrections
(at the supersymmetric scale set by the mass of the top-squarks) 
and require a genuine
two-loop calculation; they simply cannot be obtained  from RG arguments.

The status of these higher-loop calculations of the radiatively corrected 
$m_{h^0}^2$ is the following. Higher-order logarithmic corrections were 
included in studies which used RG techniques almost since the
dramatic impact of radiative corrections on $m_{h^0}$ was first
recognized. Hempfling and Hoang \cite{H2} were the first to perform a
genuine two-loop calculation of $m_{h^0}$ which also included
non-logarithmic terms. 
They computed the dominant two-loop radiative corrections [to ${\cal 
O}(\alpha_s\alpha_t)$ and ${\cal O}(\alpha_t^2)$] in the case
$\tan\beta\gg
1$ and zero top-squark mixing. Their computation also included 
the most important logarithmic corrections, which could be alternatively
incorporated by RG
resummation from one-loop results, as done {\it e.g.} in Ref.~\cite{CEQR}.
In this last paper it was also pointed out that by a judicious choice of
the
renormalization scale at which to evaluate one-loop corrections, the
higher order logarithmic corrections could be automatically taken into
account. A similar idea was later implemented in \cite{CEQW,H3} to
write down simple analytical approximations for the radiatively 
corrected $m_{h^0}^2$, obtained by iterative integration of RG equations.

Besides being limited to a particularly simple value of
$\tan\beta$, the calculation in Ref.~\cite{H2} missed the sizable
impact
of non-zero top-squark mixing in two-loop effects, that is, higher
order corrections to the contribution written 
down in Eq.~(\ref{threshold1}).
Such corrections were first
included to order ${\cal O}(\alpha_s\alpha_t)$ in the diagrammatic
calculation \cite{H2W}, and by the effective potential
method in Ref.~\cite{RenJie}.
The effect of these corrections is to shift the values of $\tAt$ that give
maximal mixing, change the corresponding Higgs mass by up to
$\sim -10$ GeV\footnote{This correction is relative to the
one-loop mass using on-shell parameters. 
The size of the correction would be much smaller if running parameters are
used in the one-loop formula (\ref{leading}).}
and introduce an asymmetry in the dependence of
$m_{h^0}$ with the sign of $\tAt$. This two-loop top-squark
mixing dependent correction was
also explicitly isolated recently by the present authors in
Ref.~\cite{ez}, which uses effective potential plus RG
techniques. Besides confirming the independent diagrammatic results of
Ref.~\cite{H2W} we 
clarified the relation of these calculations to previous
ones (in particular matching results expressed in different
renormalization schemes; see
also \cite{CH3W2}). We also derived a compact formula for the Higgs mass
(in the spirit of \cite{CEQW,H3}) which
took into account the most important radiative corrections, and
used RG techniques to include in a compact way two-loop
LL and NTLL corrections.
With the ${\cal O}(\alpha_s\alpha_t)$ radiative
corrections organized in this way, 
we find that the mixing-dependent genuine two-loop threshold corrections 
are generally small ($\simlt 3$ GeV).

Nevertheless, the large computing effort just reviewed did not exhaust the
potentially important radiative corrections: the two-loop
${\cal O}(\alpha_t^2)$ top-squark-mixing-dependent 
corrections to $m_{h^0}$ remained unknown
to this day, while it is clear that they could compete in principle with
the  ${\cal O}(\alpha_s\alpha_t)$ contributions. The purpose of this paper
is to complete the calculation performed in \cite{H2,RenJie,ez} by using
effective potential techniques (plus RG techniques) to compute
such ${\cal O}(\alpha_t^2)$ contributions for general
top-squark mixing parameters and any value of $\tan\beta$. 
The results in this paper can be considered the most 
complete and accurate approximation to
$m_{h^0}$ presented in the literature. 

The structure of the paper is the following: the next Section describes
the strategy of our calculation and presents some analytical formulae for
$m_{h^0}$, obtained in the limit of $m_{\tilde{t}}\gg m_t$. Section~3 goes
one step ahead implementing the RG-improvement of such
approximations and, in doing so, clarifies the
organization of the higher order radiative corrections. 
This procedure is not only important to provide a clearer physical
picture in connection with the effective field theory
but also to classify those corrections calculated in Sec. 2 
into a numerically dominant and compact part plus smaller finite threshold 
correction terms.  In Section 4 we present our numerical results for the
Higgs mass, illustrate the size of the new corrections and check the
validity of our analytical approximation formulae. We draw some
conclusions in Section 5.

Several appendices are devoted to technical details of different
aspects of the calculation. Appendix~D is worth special mention as it 
contains the two-loop ${\cal O}(\alpha_t^2)$ MSSM effective potential used
as starting point of our
calculation and first computed in this paper.

\section{${\cal CP}$-even Higgs boson masses to two-loop order}

The momentum-dependent mass-squared matrix for the ${\cal CP}$-even Higgs
bosons of the MSSM in the interaction eigenstate basis $h_1,~h_2$ is
\begin{equation}
{\cal M}^2_h(p^2)\ =\ \left[
\begin{array}{cc}
m^2_Z c_\beta^2 + m^2_{A^0} s^2_\beta + \Delta{\cal M}^2_{11}(p^2) &
-(m^2_Z + m^2_{A^0}) s_\beta c_\beta + \Delta{\cal M}^2_{12}(p^2) \\
-(m^2_Z + m^2_{A^0}) s_\beta c_\beta + \Delta{\cal M}^2_{21}(p^2) &
m^2_Z s_\beta^2 + m^2_{A^0} c_\beta^2 + \Delta{\cal M}^2_{22}(p^2)
\end{array}\right]\ ,\label{mhmatrix}
\end{equation}
where $s_\beta\equiv\sin\beta$ and $c_\beta\equiv\cos\beta$.
The mass parameters $m_Z$ and $m_{A^0}$ are the (scale-dependent) running
masses of the $Z$-boson and 
${\cal CP}$-odd Higgs boson $A^0$; they are related to the on-shell
masses $M_Z$ and $M_{A^0}$ (we use capital letters to distinguish on-shell
parameters from running ones) by
\begin{equation}
m_Z^2\ =\ M^2_Z+{\rm Re}\ \Pi^T_{ZZ}(M^2_Z)\ ,\qquad
m^2_{A^0}\ =\ M^2_{A^0}+{\rm Re}\ \Pi_{AA}(M^2_{A^0})
-s_\beta^2 {T_1\over v_1}-c_\beta^2 {T_2\over v_2}\ ,\label{ZH}
\end{equation}
where $\Pi^T_{ZZ}$ is the transverse part of the $Z$-boson self-energy
and $\Pi_{AA}$ the $A^0$-boson self-energy, $T_1,~T_2$ are the tadpoles
of the ${\cal CP}$-even (real) fields $h_1,~h_2$. Their explicit one-loop
expressions can be found {\it e.g.} in Ref.~\cite{PBMZ}.

In (\ref{mhmatrix}),
$\Delta{\cal M}^2$ stands for the contributions from 
radiative corrections. They are
\begin{equation}
\Delta{\cal M}^2_{ij}(p^2)
\ =\ -\Pi_{i j}(p^2)+{T_i\over v_i}\delta_{ij}\ ,
\qquad i,j=1,2\ ,\label{HH}
\end{equation}
where $\Pi_{ij}$ is the self-energy matrix of the Higgs fields $h_1$ and
$h_2$. The masses, $m_{h^0}$, $m_{H^0}$,
of the two ${\cal CP}$-even Higgs bosons
are then 
obtained from the real part of the poles of the propagator matrix,
\begin{equation}
{\rm Det}\biggl[m_{h^0, H^0}^2{\bf 1}
-{\cal M}^2_h(m_{h^0, H^0}^2)\biggr]\ =\ 0\ .
\label{det}
\end{equation}
The radiatively corrected mixing angle $\alpha$
is obtained as the angle of that rotation which diagonalizes ${\cal
M}_h^2$ (for some choice of $p^2$, say $p^2=m_{h^0}^2$): 
\be
\tan 2\alpha=
\frac{2({\cal M}_h^2)_{12}}{({\cal M}^2_h)_{11}-({\cal M}^2_h)_{22}}\ .
\ee
Computing Higgs boson masses to a certain order of perturbation
theory then requires calculating the self-energies and tadpoles in
Eqs.~(\ref{ZH}) and (\ref{HH}) to that order. 

In the effective potential approach \cite{radEP,CEQR,H2,RenJie,ez}, 
self-energies and tadpoles can be calculated as derivatives of the Higgs
potential $V$ according to:
\begin{equation}
T_i\ =\ -
\left.\left[{\partial V(h_1,h_2)\over\partial h_i}
\right]\right|_{h_1=v_1,h_2=v_2}\ ,\qquad
\Pi_{ij}(0)\ =\ -
\left.\left[{\partial^2V(h_1,h_2)\over\partial h_i\partial h_j}
\right]\right|_{h_1=v_1,h_2=v_2}\ .
\label{polder}
\end{equation}
Note that the $\Pi$'s obtained from derivatives of $V$ have zero
external momentum.

In the limit $M_{A^0}\gg M_Z$, the lightest ${\cal CP}$-even Higgs state
lies along the
direction of the breaking in field space \cite{ce}, that is,
$\alpha\rightarrow\beta-\pi/2+{\cal O}(m_Z^2/m_A^2)$, and its radiatively 
corrected mass has a very simple expression 
\begin{equation}
\label{mhep}
M^2_{h^0}\ =\ {4 m^4_t\over v^2}\left({d\over d m^2_t}\right)^2 V
-{\rm Re}\ \Pi_{hh}(m^2_{h^0})+{\rm Re}\ \Pi_{hh}(0)\ ,
\end{equation}
which is exact up to corrections of order 
${\cal O}(m_Z^4/m_{A^0}^2)$.\footnote{
This formula can be proved as follows: 
If $m_{A^0}^2\gg m_Z^2$, $\alpha\rightarrow\beta-\pi/2$,
and we can therefore use the approximation $\Delta m_{h^0}^2
\simeq\Delta{\cal M}_{11}^2 c_\beta^2+\Delta{\cal M}_{22}^2 s_\beta^2
+2\Delta{\cal M}_{12}^2 s_\beta c_\beta$, up to higher order terms
in $m^4_Z/m^2_{A^0}$. Observing  that the 
potential $V$ depends on the fields $h_1$ and $h_2$ only through
(field-dependent) top quark mass 
and the off-diagonal elements of
the top-squark mass-squared matrix 
and using (\ref{polder}), we can easily express the partial derivatives
of $V$ in terms of the total derivative in (\ref{mhep}).
A similar formula was already used in \cite{H2}.}

In Eq.~(\ref{mhep}), $V$ is the projection of $V(h_1,h_2)$ along the
light Higgs $h=
h_1 c_\beta+h_2s_\beta$: $V(h)=V(h_1\rightarrow h c_\beta,h_2\rightarrow 
h s_\beta)$. Then $V(h)$ can be expressed as a function of $m_t$ using 
$h\rightarrow m_t\sqrt{2}/(h_t s_\beta)$. We decompose $V$ in its
$n^{th}$-loop 
pieces $V_n$ (explicitly given in Appendix~D) as $V=V_0+V_1+V_2$. The
tree-level part $V_0$ is the only one
in which we keep non-zero electroweak gauge couplings. We approximate the
one-loop part $V_1$ by its ${\cal O}(\alpha_t)$ piece coming
from top quark/squark loops. The two-loop part $V_2$ is approximated by
$V_{2s}+V_{2t}$, where $V_{2s}$ is the ${\cal O}(\alpha_s\alpha_t)$ part
and $V_{2t}$ the ${\cal O}(\alpha_t^2)$ one.

Next, $\Pi_{hh}(p^2)$ is the light Higgs self-energy at external momentum
$p$, related to the self-energies of $h_{1,2}$ by
\begin{equation}
\Pi_{hh}(p^2)\equiv
\Pi_{11}(p^2)\sa^2+\Pi_{22}(p^2) \ca^2
-2\Pi_{12}(p^2)\sa\ca\ .
\end{equation}
Notice that the self-energy difference in
(\ref{mhep}) involves non-zero external momentum and would require a
diagrammatic two-loop calculation. However,
throughout this paper we work in the approximation of neglecting in
the radiative corrections all couplings other than $h_t$ or $g_3$. In that
case, realizing that at tree level $m_{h^0}$ depends only on electroweak
gauge couplings while its dependence on $h_t$ appears only at one-loop, we
can write 
\be
\label{dPi}
\Pi_{hh}(m^2_{h^0})-\Pi_{hh}(0)\simeq m^2_{h^0}
\left. \frac{d}{d p^2}\Pi_{hh}(p^2)\right|_{p^2=0}
\ ,
\ee
which gives rise to ${\cal O}(\alpha_t^2)$ contributions from the one-loop
${\cal O}(\alpha_t)$ self-energy $\Pi_{hh}$.

In Section~4 we present the numerical results of such procedure for the
two-loop potential with ${\cal O}(\alpha_s\alpha_t)$ and ${\cal
O}(\alpha_t^2)$ corrections included. The general expression for the
${\cal O}(\alpha_s\alpha_t)$ potential was first computed in
\cite{RenJie} while the  complete ${\cal O}(\alpha_t^2)$ terms were still
missing. Both contributions to $V$ are given in Appendix~D. 

It is useful, both for a better understanding of the numerical results
and for practical applications, to derive an analytical expression for the
light Higgs mass in the case of a large hierarchy between the
supersymmetric scale and the electroweak scale (say when the SUSY scale is 
of order 1 TeV). Such limit is interesting because it maximizes the
radiative corrections to $m^2_{h^0}$ (so that it corresponds to the most
pessimistic scenario for Higgs searches; the case one should be able to
discard to rule out the MSSM), and at the same time simplifies the 
structure of the radiative corrections, avoiding the proliferation of
a multitude of different supersymmetric thresholds.

We consequently assume now that all supersymmetric particles have
roughly the same mass $M_S\gg M_Z$. In more detail, focusing on the
particles relevant for the radiative corrections to $m_{h^0}$, we take
equal 
soft masses $\MQ=\MU=M_S$ for the top-squarks (with diagonal masses
$m_{\tilde{t}}^2\simeq M_S^2+m_t^2$). The two eigenvalues and mixing angle
of the top-squark squared-mass matrix are then
\begin{equation}
m^2_{{\tilt}_1}\ =\ \mtilt^2+m_t\xt\ ,\qquad
m^2_{{\tilt}_2}\ =\ \mtilt^2-m_t\xt\ ,\qquad
s_t^2\ =\ c_t^2\ =\ {1\over 2}\ ,
\label{stopspectrum}
\end{equation}
We also take the same mass $M_S$ for the
gluino and the pseudoscalar Higgs [this means in particular that we can
use Eq.~(\ref{mhep}) for the light Higgs boson]. In principle we admit the
possibility that the $\mu$ parameter could be smaller than
$M_S$, in which case
we expect that one chargino and two neutralinos will have masses $\sim
|\mu|$  below the common supersymmetric threshold. In this situation, which
broadly corresponds to the case of a common heavy SUSY scale, we find
that, using the operator
\be
{\cal D}^2_{m}\equiv{4 m^4_t\over v^2}\left({d\over d m^2_t}\right)^2,
\ee
the different parts entering (\ref{mhep}) are 
\begin{eqnarray}
\label{mhder0}
{\cal D}^2_{m} V_0
&=& m_Z^2\cos^22\beta,\\
\label{mhder1}
{\cal D}^2_{m} V_1
&=& {3m^4_t\over2\pi^2 v^2}\biggl(\Lst
+\hxt^2-\frac{\hxt^4}{12}\biggr),\\
\label{mhder2s}
{\cal D}^2_{m}  V_{2s}
&=& {\alpha_sm^4_t\over\pi^3 v^2}
\Biggl\{\LLst-2\LLs+2\LLt
+\Lt -1+\biggl(-1+2\Ls+2\Lt\biggr)\hxt\nonumber\\
&&\hspace{1.2cm}+
\biggl(1-2\Ls\biggr)\biggl(\hxt^2+\frac{\hxt^3}{3}\biggr)
-{\hxt^4\over12}\Biggr\},\\
{\cal D}^2_{m} V_{2t}
&=& {3\alpha_t m_t^4\over16\pi^3 v^2}\left\{
9\LLs-6\Lt\Ls-3\LLt+2[3f_2(\hmu)-3f_1(\hmu)-8]\Lst\right.\nonumber\\
&+&6\hmu^2\biggl(1-\Ls \biggr)
-2(4+\hmu^2)f_1(\hmu)+4f_3(\hmu)-\frac{\pi^2}{3}\nn\\
&+&\left[(33+6\hmu^2)\Ls-10-6\hmu^2-4f_2(\hmu)+(4-6\hmu^2)f_1(\hmu)
\right]\hxt^2\nonumber\\
&+&\left[-4(7+\hmu^2)\Ls+23+4\hmu^2+2f_2(\hmu)-2
(1-2\hmu^2)f_1(\hmu)\right]\frac{\hxt^4}{4}
\nonumber\\
&+&\frac{1}{2}s_\beta^2\hxt^6\biggl(\Ls-1\biggr)+
c_\beta^2\biggl[3\ln^2{M_S^2\over m_t^2}  
+7 \ls-4\lt-3+60K+\frac{4\pi^2}{3}\nonumber\\
&+&\biggl(12-24K-18\ls\biggr)\hxt^2
-\biggl(3+16K-3\ls\biggr)(4\hxt\hyt+\hyt^2)\nn\\
&+&\biggl(-6+{11\over2}\ls\biggr)\hxt^4
+\biggl(4+16K-2\ls\biggr)\hxt^3\hyt \nonumber \\ 
&+&\left(\frac{14}{3}+24K-3\ls\right)\hxt^2\hyt^2
-\left(\frac{19}{12}+8K-{1\over2}\ls\right)\hxt^4\hyt^2\biggr]\Biggr\}\ ,
\label{mhder2t}
\end{eqnarray}
The notations used are 
$\hxt=\hAt+\hmu\cot\beta$,
$\hyt=\hAt-\hmu\tan\beta$, 
with reduced parameters $\hat{z}\equiv z/M_S$, 
and (see Appendix~A) $K\simeq  -0.1953256$. 
We also use the following non-singular functions of $\hmu$
\bear
f_1(\hmu)&=&\frac{\hmu^2}{1-\hmu^2}\ln\hmu^2,\nonumber\\
f_2(\hmu)&=&\frac{1}{1-\hmu^2}\left[1+\frac{\hmu^2}{1-\hmu^2}\ln\hmu^2
\right],\nonumber\\
f_3(\hmu)&=&\frac{(-1+2\hmu^2+2\hmu^4)}{(1-\hmu^2)^2}
\left[\ln\hmu^2\ln(1-\hmu^2)
+Li_2(\hmu^2)-\frac{\pi^2}{6}-\hmu^2\ln\hmu^2
\right],
\eear
with $f_1(0)=0$, $f_2(0)=1$, $f_3(0)=-\pi^2/6$  and 
$f_1(1)=-1$, $f_2(1)=1/2$, $f_3(1)=-9/4$.

Finally,  the correction for non-zero external
momentum in Eq.~(\ref{mhep}) is given by (see Appendix~C)
\begin{equation}
{\rm Re}~\biggl[-\Pi_{hh}(m^2_{h^0})+\Pi_{hh}(0)\biggr]
\ =\ {h_t^2\over\olf} 
m_{h^0}^2 s^2_\beta \biggl(3\lntq+2-{\hxt^2\over2}\biggr)\ .
\label{dpi}
\end{equation}

The parameters that appear in these expressions are running parameters,
evaluated in the $\dr$-scheme and satisfy MSSM RG equations. In fact, it
can be checked that the (physical) Higgs mass, given by Eq.~(\ref{mhep}),
is renormalization-scale independent (up to two-loop order),
as it should. This scale independence
is at the root of the RG-resummation procedure discussed in the next
section. It is evident that for different values of the renormalization
scale, the magnitude of the two-loop corrections will change, so that it
should be possible to choose the scale in such a way that the bulk of the
corrections is transferred to the one-loop terms (which depend 
on the scale implicitly). 

Therefore, the magnitude and relevance of the two-loop corrections depends 
on the definition of the mass parameters that enter the one-loop
corrections. It is in this respect convenient to write down the two-loop
expressions just obtained in the particular case in which all mass
parameters in the one-loop correction are the OS ones. This is also
useful to compare with explicit diagrammatic
calculations. The relationships between running and OS parameters are
listed in Appendix~C. Using them, we obtain for the two-loop correction to
$m_{h^0}^2$:
\begin{eqnarray}
\Delta m_{h^0}^2 &=& 
\frac{\alpha_s m_t^4}{\pi^3v^2}\Biggl\{
-3\LLst-6\lst+6\hxt-3\Lst~\hxt^2-\frac{3}{4}\hxt^4\Biggr\}\nn\\
&+&{3\alpha_t m_t^4\over 16\pi^3 v^2}\Biggl\{
\biggl(3 \LLst + 13 \Lst\biggr)s_\beta^2 -\frac{7}{2}
-{\pi^2\over3}-3\hmu^2
-(11-\hmu^2+3\hmu^4)f_1(\hmu)\nonumber\\
&&\hspace{1.2cm}-3(1-\hmu^2)^2\ln(1-\hmu^2)
+3f_2(\hmu) +4f_3(\hmu)
+c_\beta^2\biggl(
60 K + {13\over2} + {4\pi^2\over3}\biggr)\nn\\
&&\hspace{1.2cm}+\biggl[3 s^2_\beta\Lst
+\frac{73}{2}+9\hmu^2+f_1(\hmu)-7f_2(\hmu)
-c^2_\beta\biggl({69\over2}+24K\biggr) 
\biggr] \hxt^2\nn\\
&&\hspace{1.2cm}+\frac{1}{6}
\biggl[-26-9\hmu^2+3f_1(\hmu)+3f_2(\hmu)
+{61\over2}c^2_\beta\biggr]\hxt^4
+{s^2_\beta\over2}\hxt^6\nn\\
&&\hspace{1.2cm}+3(1-\hmu^2)\left[(2-3\hmu^2)f_1(\hmu)-
(1+3\hmu^2)\ln(1-\hmu^2)
\right]\biggl(\hxt^2-\frac{\hxt^4}{6}
\biggr)\nn\\
&&\hspace{1.2cm}+c^2_{\beta}\biggl[(3-16 K-\pi\sqrt{3})(4\hxt\hyt+\hyt^2)
+\biggl(16K+{2\pi\over\sqrt{3}}\biggr)\hxt^3\hyt\nn\\
&&\hspace{1.2cm}+\biggl(-{4\over3}+24 K+\pi\sqrt{3}\biggr)\hxt^2\hyt^2
-\biggl({7\over12}+8 K+{\pi\over2\sqrt{3}}\biggr)\hxt^4\hyt^2\biggr]\nn\\
&&\hspace{1.2cm}+\biggl(2\hxt-{\hxt^3\over3}\biggr)
\biggl[\biggl(-3+{2\pi\over\sqrt{3}}\biggr) c^2_\beta \hxt\hyt^2
-\biggl(3\ln{m_t X_t\over M_S^2} + \ln4\biggr) s^2_\beta \hxt^3\biggr]
\Biggr\}
\label{mhOS}
\end{eqnarray}
We emphasize that this expression gives the two-loop corrections when the 
one-loop contribution (\ref{mhder1}) is expressed in terms of OS
parameters, that is,
\be
\left[\Delta
m_{h^0}^2\right]_{1-loop}^{\rm OS}=\frac{3g^2M_t^4}{8\pi^2M_W^2}
\left[\ln\frac{M_{\tilde{t}}^2}{M_t^2}+\left(\frac{X_t^{\rm 
OS}}{M_{\tilde{t}}}
\right)^2-\frac{1}{12}\left(\frac{X_t^{\rm OS}}{M_{\tilde{t}}}
\right)^4\right].\label{mhOS1l}
\ee

Several features of Eq.~(\ref{mhOS}) are worth commenting.
First, if we restrict Eq.~(\ref{mhOS}) to $\tan\beta\gg 1$ and zero
$A_t$, to compare with the result of Ref.~\cite{H2}, we find the same
logarithmic terms.
However, the  ${\cal O}(\alpha_t^2)$ finite term is different. In
particular, that term is sensitive to the value of the
parameter $\mu$,  contrary to what is stated in Ref.~\cite{H2}.
Nevertheless, the result quoted for that finite term in Ref.~\cite{H2}
is inside the range we would find by varying $\hmu^2$ from 0 to 1, and the
impact of this $\mu$-dependence on the final Higgs mass is quite small.

Second, we see that radiative corrections no longer depend on $A_t$ and
$\mu$ in the combination $X_t$ that appears through the
off-diagonal entry of the top-squark mass matrix: besides the explicit
dependence on the parameter $\mu$ already noticed, the quantity $Y_t$ also
introduces a different combination of $A_t$ and $\mu$.
This dependence on $Y_t$ originates from 
the $H-\tilde{t}-\tilde{b}$ and $H-\tilde{t}-\tilde{t}$
diagrams of Fig.~\ref{fig:Feyn}.

Third, although roughly  speaking the top-Yukawa correction has
a small pre-factor $3/16$ in comparison with the QCD correction, this does
not guarantee that the new contributions will be negligible compared to
the QCD one. In fact, we will see that for two-loop 
top-squark-mixing-dependent corrections of (\ref{mhOS}), 
the top Yukawa contributions have
opposite signs as that of the QCD corrections and could be 
as much as $60\%$ of the latter (see Fig.~\ref{fig:6}). 
In the next Section, we will follow RG methods and reorganize these corrections
in the effective theory language, with the most important corrections of
Eq.~(\ref{mhOS}) reshuffled in a RG-motivated one-loop formula.

\section{Renormalization group resummation}

Before illustrating in Section~4 the impact of the newly computed
corrections on the Higgs mass, we show in the following how the use of
renormalization group techniques \cite{radRG,CEQW,H3}  allows us to write
the previous complicated corrections [see Eq.~(\ref{mhOS})] in a simpler
and more transparent way, while at the same time it clarifies the connection to
the RG programme, which can be used to improve  the precision of the mass
formula by resummation of higher order corrections. 

We already applied this idea in Ref.~\cite{ez} to the
${\cal
O}(\alpha_s\alpha_t)$ two-loop corrections. By a convenient (and
physically well motivated) choice of the scale at which to evaluate
running parameters in the one-loop mass correction one can absorb
 large logarithms in Eq.~(\ref{mhOS}). The RG evolution of the
parameters is given by the corresponding one-loop RG functions listed in
Appendix~B.

We use the following equations to relate supersymmetric running
parameters at different scales [{\it cf.} Eqs.~(\ref{Xt}) and
(\ref{mst})]:
\be
m_{\tilde{t}}^2(Q)=m_{\tilde{t}}^2(Q')
\left\{1+\frac{1}{16\pi^2}\left[\frac{16}{3}g_3^2
-\frac{3}{2}h_t^2\left(\hxt^2s_\beta^2+\hyt^2c_\beta^2+c_\beta^2
+2-2\hmu^2\right)\right]
\ln\frac{Q'{}^2}{Q^2}\right\}\ ,
\ee
\be
X_t(Q)=X_t(Q')
-\frac{1}{16\pi^2}\left[\frac{16}{3}
g_3^2 M_S+3h_t^2(X_t + X_t s_\beta^2+Y_t c_\beta^2)\right]
\ln\frac{Q'{}^2}{Q^2}\ ,
\ee
where we have used $A_t=X_t s^2_\beta+Y_t c^2_\beta$,
$A_t^2+\mu^2=X^2_t s^2_\beta+Y^2_t c^2_\beta$ and 
$m_{H_2}^2+\mu^2=m^2_{A^0}c^2_\beta$. Notice that, to the order we work,
it is sufficient to use these one-loop LL approximations to the full RG
evolution because we are concerned with parameters that appear in a
one-loop order term.

The Standard Model $\ms$ top quark mass $\overline{m}_t$ 
and the Higgs VEV $\overline{v}$ 
are related to the on-shell mass $M_t$ and MSSM VEV $v$ by
[{\it cf.} Eqs.~(\ref{mtQ}) and (\ref{v2Q}), from which relevant
terms can be easily identified]
\be
\label{RGmt}
\overline{m}_t^2(Q)=
M_t^2\left[1-\frac{g_3^2}{6\pi^2}\left(4-3\ln\frac{m_t^2}{Q^2}\right)
+\frac{h_t^2s_\beta^2}{32\pi^2}\left(8-3\ln\frac{m_t^2}{Q^2}\right)\right]\ ,
\ee
\be
\label{RGv}
\overline{v}^2(Q)=v^2(Q)
\left[1+\frac{h_t^2s_\beta^2}{32\pi^2}\hxt^2\right]\ .
\ee
We also use one-loop LL solutions of the SM RG equations to  relate
these parameters at different scales:
\be
\overline{m}^2_t(Q)=\overline{m}^2_t(Q')
\left[1+\frac{1}{16\pi^2}\left(8g_3^2-\frac{3}{2}h_t^2s_\beta^2\right)
\ln\frac{Q'{}^2}{Q^2}\right]\ ,
\ee
\be
\overline{v}^2(Q)=\overline{v}^2(Q')
\left[1+\frac{3h_t^2s_\beta^2}{16\pi^2}\ln\frac{Q'{}^2}{Q^2}\right].
\ee

Using the above equations,
we find the following compact expression for the Higgs boson mass, which
is one of the main results of this paper
\be
M^2_{h^0}=M_Z^2\cos^22\beta+
{3\overline{m}_t^4 (Q_t)\over2\pi^2 \overline{v}^2(Q_1^*)}
\ln{m^2_{\tilt}(Q_{\tilde{t}})\over \overline{m}_t^2(Q_t')}
+\Delta_{\rm th}^{(1)} m_{h^0}^2 
+\Delta_{\rm th}^{(2)} m_{h^0}^2\ .
\label{1loop}
\ee
The one-loop threshold correction is
\be
\Delta_{\rm th}^{(1)} m_{h^0}^2
={3\overline{m}^4_t(Q_{\rm th})\over2\pi^2
\overline{v}^2(Q_2^*)}
\biggl[\hxt^2(Q_{\rm th})-\frac{\hxt^4(Q_{\rm th})}{12}\biggr]\ ,
\label{thr1}
\ee
and the two-loop threshold correction reads
\begin{eqnarray}
\Delta_{\rm th}^{(2)} m_{h^0}^2 &=& 
\frac{\alpha_s m_t^4}{\pi^3v^2}\left[
-2\hxt
-\hxt^2
+\frac{7}{3}\hxt^3
+\frac{1}{12}\hxt^4
-\frac{1}{6}\hxt^5
\right]\nn\\
&+&{3\alpha_t m_t^4\over 16\pi^3 v^2}\Biggl\{
R_0(\hmu)+R_2(\hmu)\hxt^2+R_4(\hmu)\hxt^4
-\frac{1}{2}s_\beta^2\hxt^6\nn\\
&+&c^2_{\beta}\biggl[60 K - {9\over2} + {4\pi^2\over3}
-(3+16 K)(4\hxt\hyt+\hyt^2)+ (15-24K) \hxt^2\nn\\
&-&{25\over4}\hxt^4+(4+16K)\hxt^3\hyt+\biggl({14\over3}+24
K\biggr)\hxt^2\hyt^2
-\biggl({19\over12}+8K\biggr)\hxt^4\hyt^2\biggr]\Biggr\}.
\label{thr2}
\end{eqnarray}
We have used the short-hand notation
\begin{eqnarray}
R_0(\hmu)&=&-\frac{9}{2} -{\pi^2\over3}
+6\hmu^2
-(11+2\hmu^2)f_1(\hmu)+3f_2(\hmu)+4f_3(\hmu),\nn\\
R_2(\hmu)&=&-11-\hmu^2[6+6f_1(\hmu)+10f_2(\hmu)],\nn\\
R_4(\hmu)&=&6+\hmu^2[1+f_1(\hmu)+f_2(\hmu)]\ .
\end{eqnarray}

The scales required in (\ref{1loop},\ref{thr1}) are
\[
Q_t\ =\ \sqrt{m_t\mtilt}\ ,\qquad Q_t'\ =\ (m_t\mtilt^2)^{1/3}\ ,
\qquad Q_{\tilde{t}} \ =\ Q_{\rm th}\ =\ \mtilt\ ,\nn
\] 
\begin{equation}
\label{scales}
Q_1^*\ =\ e^{-1/3}m_t\simeq 0.7 m_t\ ,\qquad \qquad
Q_2^*\ =\ e^{1/3}m_t\simeq 1.4 m_t\ .
\end{equation} 
It is a non-trivial check of our calculation that the values of the scales
(\ref{scales}) required to re-absorb the large $\ln(M_S^2/m_t^2)$ 
logarithms in the two-loop corrections
are consistent with the ones obtained in \cite{ez} for the QCD corrections
alone. We see, in particular, that the uncertainty found there in the
definition of the scales $Q_t'$ and $Q_{\tilde{t}}$ is here resolved by
the need of absorbing the new radiative corrections.

We still find a somewhat complicated expression for the threshold
correction $\Delta_{\rm th}^{(2)} m_{h^0}^2$, due to the fact that we
have kept
free the $\mu$ parameter. Expressions for the two limiting cases of heavy
$\mu$
($\mu\simeq M_S$) and light $\mu$ ($\mu\ll M_s$) can be readily derived.
In both cases, the resulting threshold correction is much simpler
than the general case (\ref{thr2}) and contains no more logarithms.
Explicitly, for $\mu\ll M_s$ we find
\be
R_0(0)\ = \  \frac{\pi^2}{3}-\frac{3}{2},\qquad
R_2(0)\ = \  -11,\qquad
R_4(0)\ = \  6,
\ee
and for $\mu\simeq M_S$:
\be
R_0(1)\ = \  7-\frac{\pi^2}{3},\qquad
R_2(1)\ = \  -16,\qquad
R_4(1)\ = \ \frac{13}{2}  .
\ee

It is perhaps convenient to make more explicit the connection between our
results and those obtained in the RG approach 
(see, {\it e.g.}, Ref.~\cite{H3}). 
To be concrete, let us
assume $|\mu|=M_S$ so that all supersymmetric particles (including charginos
and neutralinos) have masses of order $M_S$; below that scale, the
effective theory is the SM. 
The light Higgs quartic coupling
$\lambda$ at $M_S$ consists of a 
tree-level part plus higher-order threshold corrections which arise from
the heavy decoupling supersymmetric particles, it can be evolved
down to the electroweak scale, say $Q=m_t$, using the SM RGEs;  
at that scale $\lambda$ is related to the physical Higgs mass. 
This procedure should reproduce all the
logarithmic corrections we have found. 

More explicitly, defining $\bl\equiv d\lambda/d\ln Q^2$, we can write
\be
\label{RGformula}
\lambda(\mt)=\lambda(\mst)-\bigint_{Q=\mt}^{\mst} \bl\ d\ln Q^2.
\ee
We use a special notation for the high and low scales between which we run
$\lambda$ to distinguish them from other definitions of $m_t$ and
$m_{\tilde{t}}$ that appear in the paper. These quantities are defined by:
\be
\mt\equiv \overline{m}_t(\mt)\ ,\qquad
\mst\equiv m_{\tilde{t}}(\mst)\ ,
\ee
{\it i.e.}, they are the running masses evaluated at a scale equal
to the corresponding mass. This is the natural definition in the RG
approach.

Making a loop expansion of $\bl$ in (\ref{RGformula}) and a further
expansion around a particular value of $Q$ (say the low energy limit of
the running interval, $\mt$), we obtain to the two-loop order
\be
\label{lqt}
\lambda(\mt)\simeq\lambda(\mst)-\left[\bl^{(1)}(\mt)+\bl^{(2)}(\mt)\right]
\ln\frac{\mst^2}{\mt^2}-\frac{1}{2}\frac{d\bl^{(1)}}{d\ln
Q^2}(\mt)\ln^2\frac{\mst^2}{\mt^2}+...
\ee
where the one- and two-loop contributions to $\beta_\lambda$ are 
approximated by [neglected all couplings other than the
strong gauge coupling
$g_3$ and the SM top Yukawa coupling $g_t$ ($\equiv h_t s_\beta$)]
\bear
\bl^{(1)}&=&\frac{3 g^2_t}{8\pi^2}(-g_t^2+\lambda),\nn\\
\bl^{(2)}&=&\frac{2 g^4_t}{(16\pi^2)^2}(15 g_t^2-16 g_3^2).
\eear
We note that for a
correct
two-loop computation it is necessary to retain also the $\lambda g_t^2$
term in
$\bl^{(1)}$ because $\lambda$ gets one-loop contributions proportional to 
$g_t^4$.
$d\beta^{(1)}_\lambda/d\ln Q^2$ can be calculated from
the one-loop RG evolution of $g_t$
\be
\frac{d g_t^2}{d\ln Q^2}=\frac{g_t^2}{32\pi^2}(9g_t^2-16g_3^2).
\ee

Once $\lambda(\mt)$ is obtained from (\ref{lqt}), we extract 
the physical Higgs mass using the SM relation \cite{SZ}:
\be
\label{sz}
\lambda(\mt)\overline{v}^2(\mt)=
M_{h^0}^2\left(1-\frac{g_t^2}{8\pi^2}\right).
\ee
This correction arises from wave-function renormalization and takes into
account the fact that the physical mass is defined on-shell, and not at
zero external momentum. Its physical content is therefore similar to
the correction (\ref{dPi}) in our effective potential approach.

According to (\ref{lqt}), the large LL and NTLL 
corrections to $M^2_{h^0}$ arise solely from $\lambda(\mt)$. Additional
radiative contributions in (\ref{sz}), coming from $\overline{v}^2(\mt)$
and the wave-function correction factor, affect the large logarithmic
terms only through multiplication of $\lambda(\mt)$. It is therefore clear
that it is enough for our purposes to know these correction factors at
one-loop order. Based on this observation, we can combine both factors
together using (\ref{RGv}) to write the simpler formula
\be
\label{mhRG}
M_{h^0}^2=\lambda(\mt)\overline{v}^2({\cal Q}_1^*),
\ee
with ${\cal Q}_1^*=e^{-1/3}\mt$, in accordance with (\ref{scales}). 

It is now straightforward to show perfect agreement of $M_{h^0}$ as obtained
from the above expression with our results
(\ref{1loop}-\ref{thr2}). All logarithmic corrections up to two-loops are 
exactly reproduced while the finite part agrees if one uses as boundary 
condition at the SUSY scale 
\be
\label{lst}
\lambda(\mst)=\frac{1}{4}(g_1^2+g_2^2)\cos^22\beta
+\delta_{\rm th}^{(1)}\lambda
+\delta_{\rm th}^{(2)}\lambda,
\ee
with 
\bear
\delta_{\rm th}^{(1)}\lambda&=&
\frac{3g_t^4(Q_{\rm th})}{8\pi^2}\left[\hxt^2(Q_{\rm th})
-\frac{\hxt^4(Q_{\rm th})}{12}\right],\nn\\
\delta_{\rm th}^{(2)}\lambda&=&\frac{\Delta_{\rm
th}^{(2)}m_{h^0}^2}{v^2}.
\eear

To summarize, we find full agreement between our approximate formula 
(\ref{1loop}) for the Higgs boson mass and the RG-improved mass calculated
in the RG (or effective theory) approach, to two-loop order. The
connection to the effective theory language clarifies the 
origin of the different terms in (\ref{1loop}), and rewrites them in a
very convenient way, absorbing the large (logarithmic) two-loop effects
in the one-loop correction and leaving behind two-loop threshold
corrections which are numerically small, as we will see in the next
Section. Note that this applies in particular to the sizable  
top-squark-mixing-dependent corrections of Eq.~(\ref{mhOS}), the bulk of
which is transferred to the RG-reshuffled one-loop threshold correction of 
Eq.~(\ref{thr1}).

Knowing the boundary condition, $\lambda(\mst)$, one can
integrate (\ref{RGformula}) numerically by solving a coupled set of
differential equations (describing the two-loop evolution of $\lambda$,
$g_3$, $g_t$), find $\lambda(\mt)$ and use (\ref{mhRG}) to get the
Higgs mass. The final result will be the full RG-improved value
of $M_{h^0}$ and will resum LL and NTLL
corrections to {\em all loops} [numerical integration includes 
all the terms from the expansion around $\mt$ which were neglected in
(\ref{lqt})]. In this respect, note that our
compact formula, Eq.~(\ref{1loop}), which has been found by requiring that
logarithmic contributions are correctly
reproduced up to two-loops only, contains in fact logarithmic corrections
of higher order. It can be shown that these higher order logarithmic
corrections do not match exactly the correct ones (obtained by the RG
method) if we use simple one-loop approximations [like those given in
Eqs.~(\ref{RGmt},\ref{RGv})] to
evaluate the
parameters in (\ref{1loop}) at their corresponding scales. However,  
evaluation of those parameters by means of a full numerical integration
[similar to that in 
Eq.~(\ref{RGformula}) for $\lambda$] would correctly take into account the
LL (but not the NTLL) terms to all loops. Nevertheless, as we will see in
the next Section, the error made in neglecting logarithmic corrections of
higher order is very small for SUSY scales of interest [below $M_S\sim
{\cal O}(1)$ TeV]. If $M_S$ turns out to be significantly larger than
that (starting to be in conflict with naturalness criteria), then one
should revert to the numerical RG integration of $\lambda$ to get a
reliable estimate of the Higgs mass. Our results for the boundary
condition $\lambda(\mst)$ will still be useful in such a case.

\section{Numerical results}

\begin{figure}[p]
\postscript{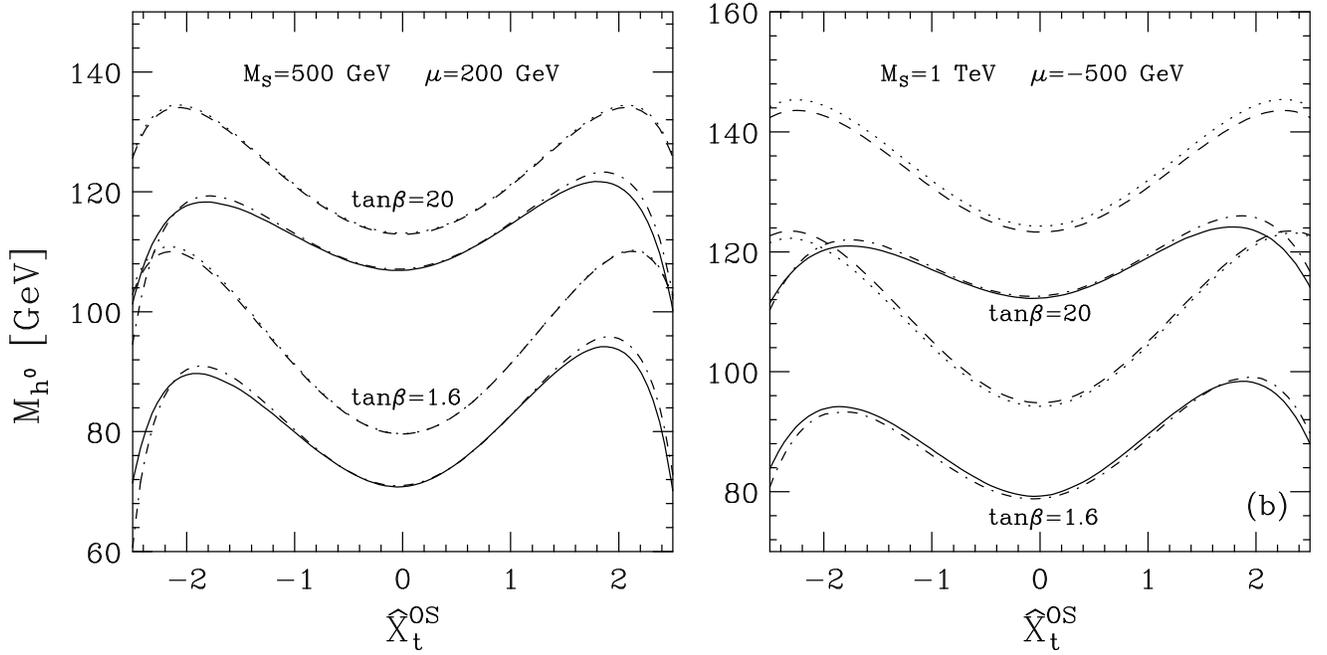}{1}
\caption{Higgs boson mass $M_{h^0}$ 
vs. the on-shell top-squark mixing parameter 
$\hat{X}^{\rm OS}_t$. Dotted, dot-dashed lines show one-loop 
and two-loop ${\cal O}(\alpha_s\alpha_t)$ 
results from the program {\tt FeynHiggs}, 
corresponding results from our numerical analyses are shown 
in dashed and solid lines respectively.}  
\label{fig:1}
\end{figure}

\begin{figure}[p]
\postscript{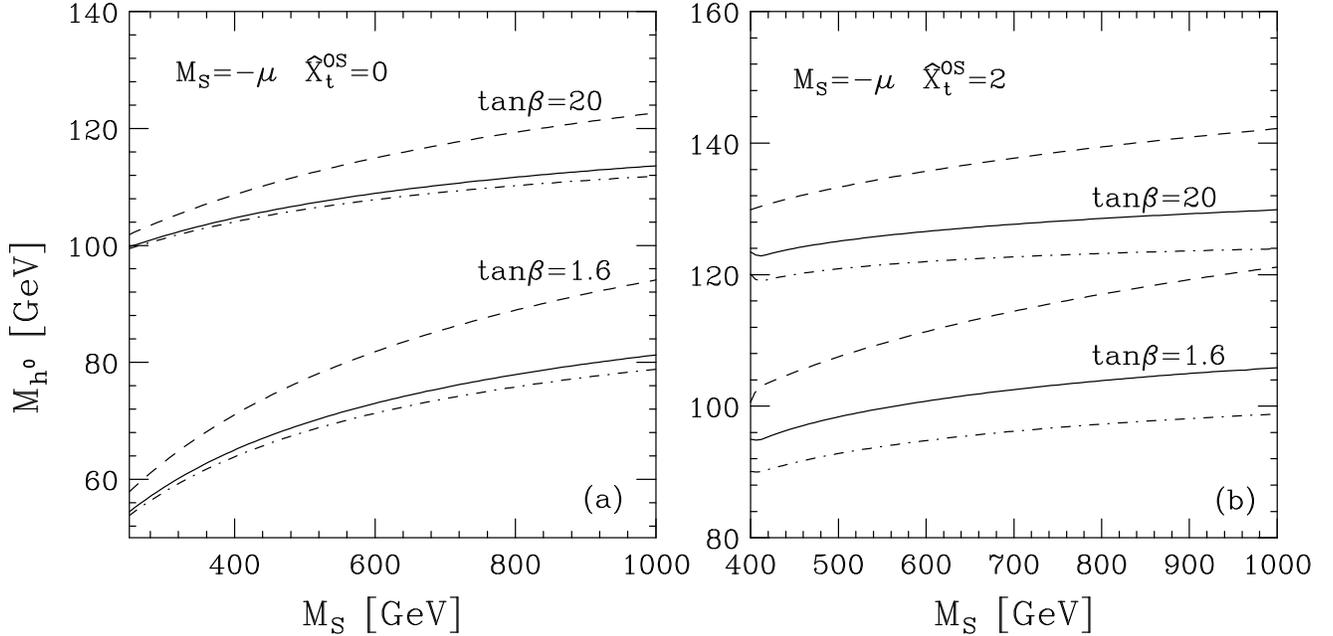}{1}
\caption{Higgs boson mass $M_{h^0}$ vs. the on-shell SUSY scale $M_S$,
for two top-squark mixing parameters $\hat{X}^{\rm OS}_t=0$ and 2. 
One-loop mass, two-loop masses to ${\cal O}(\alpha_s\alpha_t)$
and ${\cal O}(\alpha_s\alpha_t+\alpha_t^2)$ are shown 
in dashed, dot-dashed and solid lines respectively. }
\label{fig:2}
\end{figure}

\begin{figure}[tbh]
\postscript{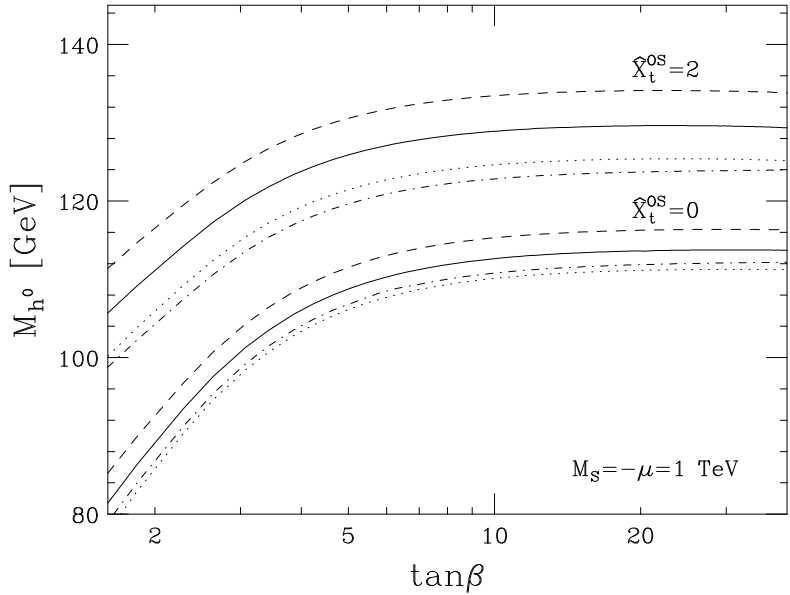}{0.6}
\caption{Higgs boson mass $M_{h^0}$ vs. 
$\tan\beta$ for the top-squark
mixing parameters $\hat{X}^{\rm OS}_t=0$ and $2$. Dot-dashed and solid
lines
correspond to two-loop Higgs boson masses to ${\cal O}(\alpha_s\alpha_t)$
and ${\cal O}(\alpha_s\alpha_t+\alpha_t^2)$ respectively for $M_t=175$
GeV.
Two-loop masses to ${\cal O}(\alpha_s\alpha_t+\alpha_t^2)$ for 
$M_t=170$ and $180$ GeV are also shown in dotted and dashed lines.}
\label{fig:3}
\end{figure}

\begin{figure}[p]
\postscript{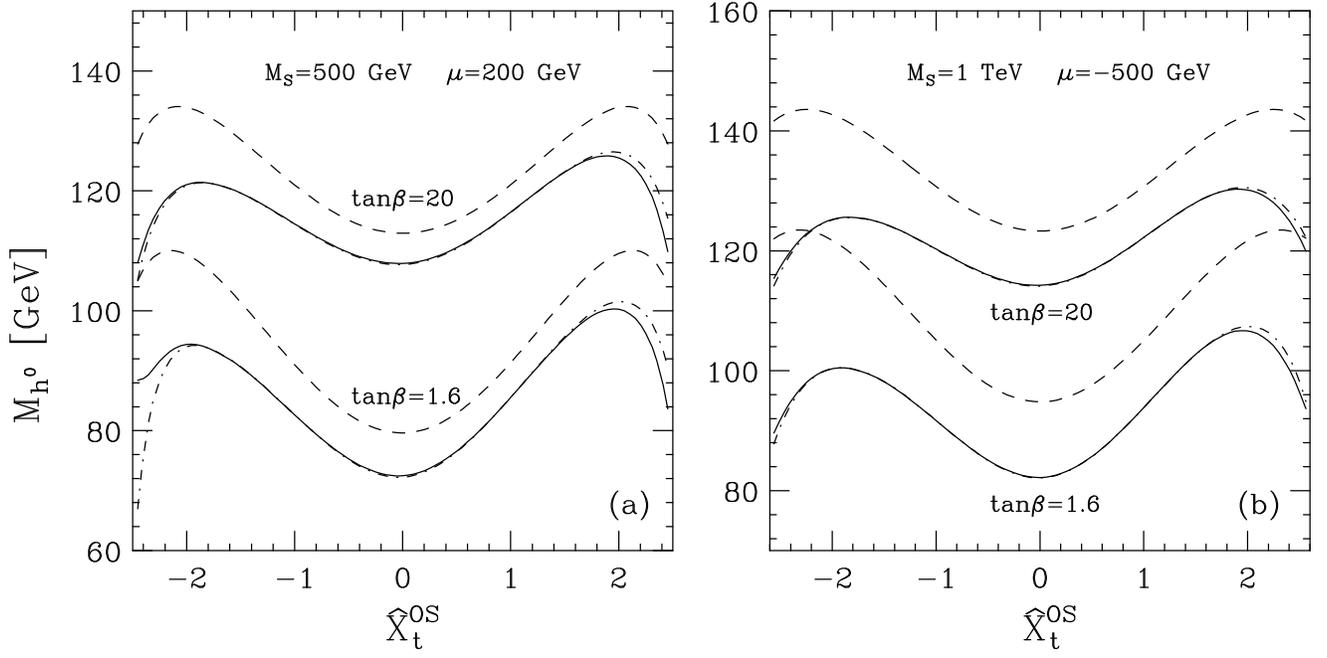}{1}
\caption{Higgs boson masses $M_{h^0}$ vs. 
$\hat{X}^{\rm OS}_t$. One-loop masses,
two-loop masses to ${\cal O}(\alpha_s\alpha_t+\alpha_t^2)$ and their
approximations
are shown in dashed, solid and dot-dashed lines. }
\label{fig:4}
\end{figure}

\begin{figure}[p]
\postscript{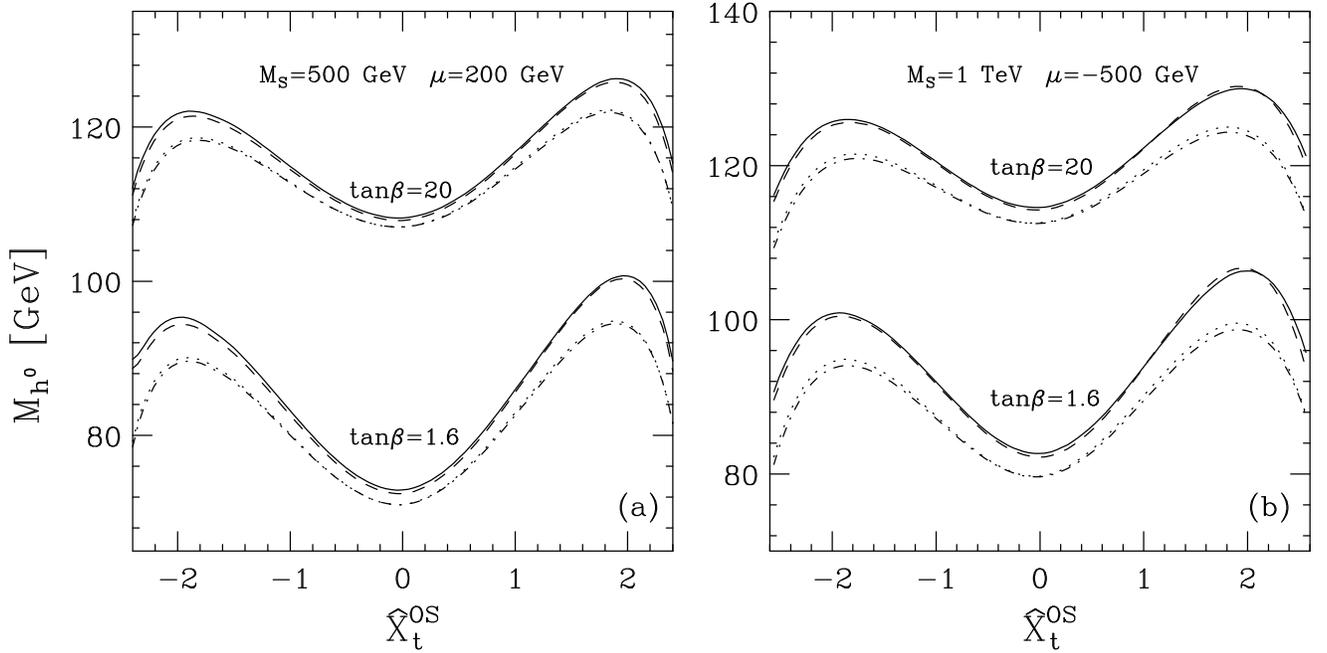}{1}
\caption{Higgs boson masses $M_{h^0}$ vs. 
$\hat{X}^{\rm OS}_t$. Two-loop masses to ${\cal O}(\alpha_s\alpha_t)$
and ${\cal O}(\alpha_s\alpha_t+\alpha_t^2)$ are shown in dotted and dashed
lines,
their corresponding RG-corrected masses are shown in dot-dashed and 
solid lines. }
\label{fig:5}
\end{figure}

\begin{figure}[p]
\postscript{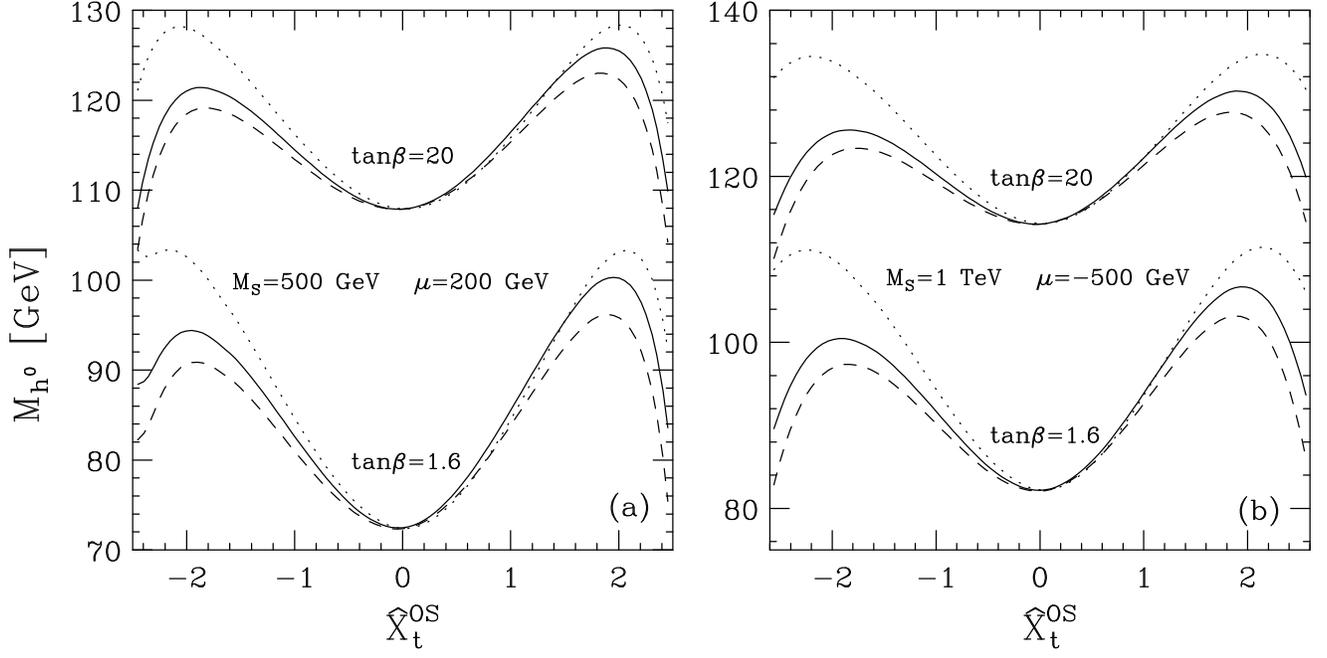}{1}
\caption{Higgs boson masses $M_{h^0}$ vs.
$\hat{X}^{\rm OS}_t$. Two-loop masses without the 
top-squark-mixing-dependent 
correction terms of Eq.~(\ref{mhOS}) are shown in dotted lines.  
The corresponding masses without the ${\cal O}({\alpha_t^2})$ 
corrections only and the full numerical results
are shown in dashed and solid lines respectively.}
\label{fig:6}

\end{figure}
\begin{figure}[p]
\postscript{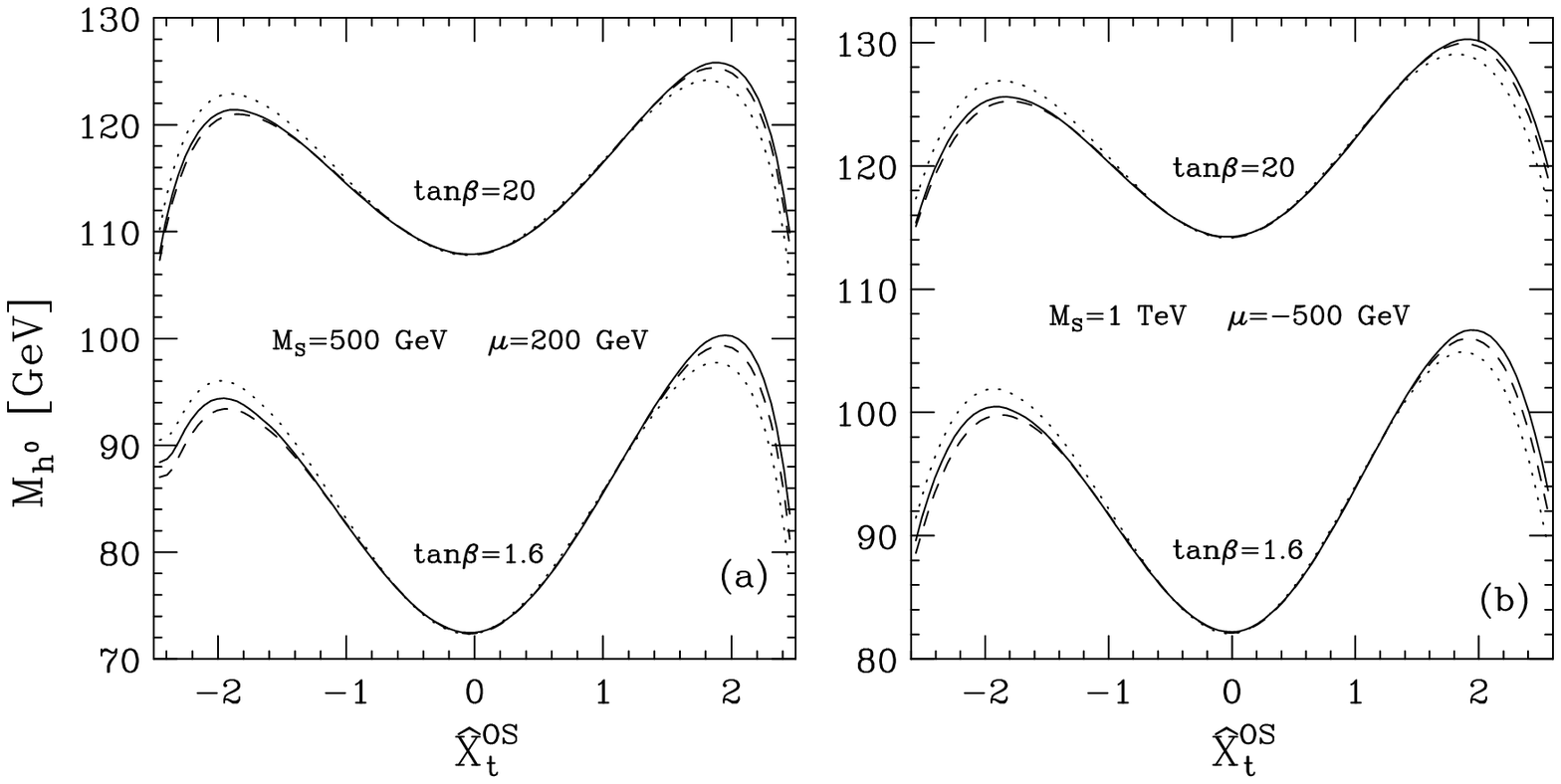}{1}
\caption{Higgs boson masses $M_{h^0}$ vs.
$\hat{X}^{\rm OS}_t$. Two-loop masses without the threshold correction
term $\Delta^{(2)}_{\rm th} m^2_{h^0}$ are shown in dotted lines.  
The corresponding masses without the ${\cal O}({\alpha_t^2})$
threshold corrections only and with the full numerical results
are shown in dashed and solid lines respectively.}
\label{fig:7}
\end{figure}

In this section we present numerical results from our two-loop study.
For the one-loop analysis  we closely follow Ref.~\cite{PBMZ}, 
which has included complete
radiative corrections from the dominant 
top quark/squark sector and the sub-dominant gauge/Higgs 
boson and neutralino/chargino sectors. 
In what follows, we shall concentrate
on two-loop radiative corrections. 

We start by sketching 
the procedure for this analysis, which is the following:
we first take as inputs the on-shell
mass parameters\footnote{For the top-squark sector, we can 
alternatively take as
inputs the on-shell top-squark masses and mixing angle.}
$M_{A^0}$, $M_t$, $\MQ^{\rm OS}$, $\MU^{\rm OS}$ and $A_t^{\rm OS}$. From
them
we can determine the values of the corresponding running parameters at
any renormalization scale $Q$. To do this, we have to calculate 
the one-loop self-energy diagrams for Higgses and top-squarks 
(the latter are collected in Appendix~C).
We also input $\tan\beta$ and $\mu$ parameters,
and convert $\alpha_s(M_Z)=0.118$ to the MSSM $\dr$ running value.
Next we calculate in the MSSM the two-loop corrections to the ${\cal
CP}$-even Higgs mass matrix, $\Delta{\cal
M}^2_{ij}$, 
from the two-loop potential (\ref{2las}) and (\ref{2lat}) using  
Eq.~(\ref{polder}). Numerically, the partial derivatives  in these
equations are
replaced by finite differences in $h_1, h_2$, {\it i.e.} we vary the 
values of these fields by a finite amount and 
recalculate the field-dependent 
top-quark mass $m^2_t = {1\over2}h_t^2 h_2^2$ and
top-squark masses $m_{\tilt_1}$, $m_{\tilt_2}$, mixing angle 
$\theta_{\tilt}$ from
Eq.~(\ref{squarkm}). With these new parameters,
the two-loop potential is reevaluated and their variations from the
reference values [calculated at $h_1^2+h_2^2=(246~{\rm GeV})^2$] are found. 
Finally, equipped with the corrections $\Delta{\cal M}^2_{ij}$,
we compute the lightest ${\cal CP}$-even Higgs boson mass by
solving Eq.~(\ref{det}).

Several approximations have been made to quantities in 
Eqs.~(\ref{mhmatrix}-\ref{HH}), in particular we neglect all dimensionless 
couplings other than the top-Yukawa coupling $h_t$ and the QCD gauge
coupling $g_3$. In this way we pick up the dominant radiative effects
only, what we term throughout leading corrections.
We notice that the two-loop self-energy of
the $Z$-boson and the
non-zero external momentum corrections to two-loop Higgs boson
self-energies
can be neglected in our calculation
since all these corrections are higher order effects in the leading
approximation. However, we need to calculate $\Pi_{AA}$ to the 
two-loop level since it has ${\cal O}(\alpha_s\alpha_t)$ and 
${\cal O}(\alpha_t^2)$ corrections and in principle could
contribute to (\ref{ZH}) at the same order as $\Delta{\cal M}^2_{ij}$. 
It is not possible to obtain these self-energies in our
current approach, and explicit two-loop calculation
of the corresponding two-point functions are needed. 
Fortunately,
the correction to $m_{h^0}$ from $\Pi_{AA}$ is always numerically negligible 
for large $m_{A^0}$ as can be easily seen from the structure of the Higgs
mass matrix (\ref{mhmatrix}).
That is, (\ref{mhep}) is correct for large $m_{A^0}$ and
we can safely neglect the $\Pi_{AA}$ corrections.
(For $m_{A^0}\sim m_Z$, a complete two-loop calculation of $m_{h^0}$ 
would need $\Pi_{AA}$.)

Fig.~\ref{fig:1} is used as calibration: we compare in it our numerical
results for $M_{h^0}$ including only up to two-loop ${\cal
O}(\alpha_s\alpha_t)$ corrections with the mass obtained by the
program {\tt FeynHiggs} \cite{FeynHiggs} which uses the explicit two-loop
diagrammatic results of Ref.~\cite{H2W}.
We choose two sets of parameters\footnote{We assume $M_3$ is positive hereafter. For a negative
$M_3$ our formulae still apply simply by simultaneous sign changes in $X_t$ and $Y_t$.}: 
(a) $M_{A^0}=M_3=\MQ^{\rm OS}=\MU^{\rm OS}=M_S=500$ GeV, $\mu=200$ GeV
and (b) $M_{A^0}=M_3=\MQ^{\rm OS}=\MU^{\rm OS}=M_S=1$ TeV, $\mu=-500$ GeV.
For each case, results for two values of $\tan\beta$ ($1.6$ and $20$)
are plotted. We find good agreement (given the fact that they
are two independent programs) between both one-loop (shown in dotted and
dashed lines) and two-loop QCD corrected (shown in dot-dashed and solid lines)
masses; this shows numerically that the two approaches are
equivalent to that order. This equivalence is easily understood
since the effective potential, 
as a generating functional \cite{cwj}, encompasses
all tadpole and self-energy diagrams (as well as 
all other multi-point functions)
which are calculated in \cite{H2W}.
The effective potential approach is more efficient for
the purpose of calculating $M_{h^0}$ and much simpler to implement
in a {\tt Fortran} program since it requires evaluating 
only one set of two-loop functions.

In Fig.~\ref{fig:2} we show the Higgs boson mass $M_{h^0}$ 
vs. the (on-shell) SUSY scale $M_S$, for two values 
of the top-squark mixing parameters $\hat{X}^{\rm OS}_t$ ($0$ and $2$).
All the physical masses $M_{A^0}$, $\MQ^{\rm OS}$, $\MU^{\rm OS}$ 
and $M_3$ have been set to $M_S$ (and the same will be done for all 
the following plots).
The dashed, dot-dashed and solid lines in this figure correspond 
to masses $M_{h^0}$ corrected to one-loop, 
two-loop ${\cal O}(\alpha_s\alpha_t)$ 
and two-loop ${\cal O}(\alpha_s\alpha_t+\alpha_t^2)$ order\footnote{In
general, we try to follow the rule that denser lines correspond to
more precise approximations.}. Fig.~\ref{fig:2}a ($\hat{X}_t^{\rm OS}=0$)
corresponds to the case of minimal left-right top-squark mixing,
and the two-loop ${\cal O}(\alpha_t^2)$ corrections are generally
small, $\lsim 2$ GeV. For Fig.~\ref{fig:2}b ($\hat{X}_t^{\rm OS}=2$), 
which roughly corresponds
to the maximal left-right top-squark mixing case, we find that the
two-loop ${\cal O}(\alpha_t^2)$ corrections are sizable ($\simeq 5$ GeV). 

In Fig.~\ref{fig:3} we examine the upper limit on the Higgs boson mass
$M_{h^0}$ by including the dominant two-loop corrections. 
We show corrected massed to the two-loop ${\cal O}(\alpha_s\alpha_t)$
and ${\cal O}(\alpha_s\alpha_t+\alpha_t^2)$ orders in dot-dashed and solid
lines for
$\hat{X}^{\rm OS}_t=0, 2$ and the top quark pole mass $M_t=175$ GeV. We
see
that maximal values for $M_{h^0}$ of $\simeq 129$ GeV can be reached 
for large $\tan\beta$ and  left-right top-squark mixing parameter
$\hat{X}^{\rm OS}_t\simeq 2$. Without two-loop ${\cal O}(\alpha_t^2)$ 
corrections, the upper bound of $M_{h^0}$ would be at $\simeq 124$ GeV.
We also show the Higgs boson masses for $M_t=180$ and 170 GeV
(including all two-loop dominant corrections)
in dashed and dot-dashed lines; the masses are increased or decreased by
$\sim$ 5 GeV respectively. We remark that 
this upper bound on $M_{h^0}$ is asymmetric
with respect to $\hat{X}^{\rm OS}_t$. For $\hat{X}^{\rm OS}_t=-2$ and 
$M_t=175$ GeV, we find the bound is about $5$ GeV lower. As is well know,
this asymmetry arises from the two-loop  ${\cal O}(\alpha_s\alpha_t)$
corrections \cite{H2W,RenJie}.

In Fig.~\ref{fig:4} we compare results from our analytical 
approximation formula 
for $M_{h^0}$ in Sec.~2 with those obtained by full numerical
evaluations. They are shown in dot-dashed and solid lines respectively.
The analytical approximation formula works remarkably well: 
it is good to a precision of $\lsim 0.5$ GeV for almost
all the parameter space. The analytical approximation has a 
complicated dependence on the $\mu$-parameter. Numerically this
dependence is quite weak: varying $\mu$ from 100 GeV to 1 TeV for
a fixed $\hat{X}_t^{\rm OS}$ changes the Higgs boson mass by less than 1
GeV.
We emphasize that the analytical formula is useful for
several reasons: (1) the logarithmic and finite corrections can be 
easily separated, and one can weight the relative importance
of these terms; (2) all terms can be traced back to the 
potential, so one can easily locate the particles giving the
biggest contributions; (3) the formula can significantly simplify
the numerical evaluations of $M_{h^0}$ to a good
precision.

In Fig.~\ref{fig:5} we further compare the results for our RG-corrected
Higgs boson masses, Eqs.~(\ref{1loop}-\ref{thr2}),
with those of the full numerical evaluation. 
For comparison, we have also shown two-loop ${\cal O}(\alpha_s\alpha_t)$
corrections and their RG-corrected results; they have been studied 
previously in \cite{ez}. As mentioned in Sect.~3, the good agreement
between these curves is an indication of the smallness of the logarithmic
corrections beyond two-loops and illustrates the accuracy of our results. 

Finally in Figs.~\ref{fig:6} and \ref{fig:7} we detail the size of the
two-loop top-squark-mixing-dependent corrections in the OS-scheme 
and their corresponding finite threshold corrections in the RG approach. 
Fig.~\ref{fig:6} shows in dotted lines
two-loop masses without including the 
top-squark-mixing-dependent corrections of Eq.~(\ref{mhOS}). 
Refs.~\cite{H2W, RenJie, ez} have already calculated
the QCD corrections, and they are depicted in dashed
lines. The difference of the solid and dashed lines is the 
two-loop ${\cal O}(\alpha_t^2)$ terms which are calculated
in this paper. We see clearly that these
terms are sizable: for large mixing parameters, they increase 
$M_{h^0}$ by about 4 GeV and $2-3$ GeV for small and large $\tan\beta$
respectively.

Fig.~\ref{fig:7} shows the effect of two-loop threshold corrections 
$\Delta^{(2)}_{\rm th} m_{h^0}^2$ evaluated following the RG-inspired
analysis of Sect.~3.
The dotted lines show the Higgs boson mass neglecting these corrections;
this would have been obtained by integrating
two-loop RG equations (with the two-loop boundary threshold
correction being set to zero), 
as we have shown in the second part of Sec.~3. 
Two-loop masses without
the ${\cal O}(\alpha_t^2)$ threshold correction and 
the complete two-loop results are shown in dashed and solid lines 
respectively. The RG reshuffling of radiative corrections has
absorbed the main part of 
the two-loop top-squark-mixing-dependent
terms of Eq.~(\ref{mhOS}) into the RG-corrected one-loop term
$\Delta^{(1)}_{\rm th} m^2_{h^0}$; the remaining genuine two-loop
threshold corrections (in the sense of the effective field theory)
are generally small, $\lsim 3$ GeV. 

\section{Conclusions}

In this paper we calculate radiative corrections to the lightest MSSM
${\cal CP}$-even Higgs boson mass to the two-loop ${\cal O}(\alpha_t^2)$
order. Our analysis extends existing two-loop diagrammatic results
\cite{H2,H2W,RenJie,ez} using a simpler effective potential approach and
provides the most complete and accurate calculation presented in the
literature. We also derive useful analytical approximation formulae,
applicable when the supersymmetric particles are heavy,
which accurately reproduce results from the full numerical study.

Our calculation includes effects which  can have an impact on the final 
Higgs mass but were neglected by previous 
studies. In particular, the two-loop ${\cal O}(\alpha_t^2)$
top-squark-mixing-dependent corrections to $M_{h^0}^2$
[see Eq.~(\ref{mhOS})] are calculated for the first term in this
paper and are numerically important.

We further simplify our analytical formula by reshuffling higher order
logarithmic corrections (using RG techniques) in a compact one-loop
expression [Eq.~(\ref{1loop})]. In
that expression all mass parameters are 
evaluated at appropriate renormalization scales chosen to 
reproduce the numerically most important leading and next-to-leading logarithmic
corrections. The remaining two-loop finite terms can be interpreted as
threshold corrections, and are numerically less important.
This RG rewriting clarifies the structure of the two-loop
corrections to $M_{h^0}^2$, identifies the most important contributions 
and links our work to the effective theory or RG approach, as we have
shown in detail in Sec.~3.

To summarize our numerical results, we have shown that two-loop top Yukawa
corrections to $M_{h^0}$ are sizable for the maximal
top-squark mixing case. They can increase the Higgs boson mass $M_{h^0}$
by as much as $5$ GeV (among which the top-squark-mixing-dependent 
corrections account for about 4 GeV) for small $\tan\beta$ where $h_t$ is
large. 
The upper bound on $M_{h^0}$ is $129\pm 5$ GeV for $M_t=175\pm 5$ GeV. 
Our final approximation 
formulae (\ref{mhOS}-\ref{mhOS1l}) and (\ref{1loop}-\ref{thr2})
have been shown to excellently
agree with the full numerical results and
can be easily implemented in precision numerical studies.

Although we have focussed in this paper on the Higgs mass, it is worth
mentioning that we have also presented in Appendix~D the MSSM two-loop
effective potential including top-quark Yukawa contributions (for general
top-squark mixing parameters and any $\tan\beta$). This knowledge may
well prove useful for other studies.

\section*{Acknowledgments}

We thank Andr\'e Hoang for correspondence.  R.-J.Z. was
supported in part by a DOE grant No. DE-FG02-95ER40896 and in part by the
Wisconsin Alumni Research Foundation.

\section*{Appendix A: One- and two-loop scalar functions}
\setcounter{equation}{0}
\renewcommand{\theequation}{A.\arabic{equation}}

\subsection*{A.1 One-loop scalar functions}

In this subsection
we define the scalar functions 
$A_0$, $B_0$, $B_1$, $B_{22}$ and $G$, which appear in one-loop
self-energy calculations.

The $A_0$ function is defined by the following momentum integral
in $d=4-2\epsilon$ dimensions
\begin{equation}
A_0(m^2)\ =\ 16\pi^2 \mu^{4-d}\int {d^dp\over i(2\pi)^d}{1\over p^2-m^2
+i\varepsilon}\ =\ 
m^2\biggl({1\over{\epsilon}}+1-\ln{m^2\over Q^2}\biggr)\ ,
\end{equation}
where $Q^2=4\pi\mu^2e^{-\gamma_E}$ is the renormalization scale,
with $\gamma_E$ the Euler constant.

The $B_0$ function is
\begin{eqnarray}
B_0(p^2,m_1^2,m_2^2) &=& 
16\pi^2 \mu^{4-d}\int {d^dq\over i(2\pi)^d}{1\over [q^2-m^2_1
+i\varepsilon][(q-p)^2-m^2_2+i\varepsilon]}\nn\\
&=&{1\over\epsilon}-\int_0^1 dx~\ln{(1-x)m^2_1+x m^2_2 - x(1-x) p^2
-i\varepsilon\over Q^2}\ .\label{B0int}
\end{eqnarray}

The remaining functions can be related to $A_0$ and $B_0$ as follows
\begin{eqnarray}
B_1(p^2,m^2_1,m^2_2) &=& {1\over 2p^2}\biggl[A_0(m^2_2)-A_0(m^2_1)+
	(p^2+m_1^2-m_2^2)B_0(p^2,m^2_1,m^2_2)\biggr]\ ,\label{B1}\\
B_{22}(p^2,m^2_1,m^2_2) &=& 
{1\over 6}\biggl[A_0(m_2^2)+2m_1^2B_0(p^2,m_1^2,m_2^2)-
(p^2+m_1^2-m_2^2)B_1(p^2,m_1^2,m_2^2)\nn\\
&&~~~+m_1^2+m_2^2-{p^2\over3}\biggr]\ ,\label{B22}\\
G(p^2,m_1^2,m_2^2) &=& (p^2-m^2_1-m^2_2)B_0(p^2,m_1^2,m_2^2)-A_0(m_1^2)
-A_0(m^2_2)\ .\label{G}
\end{eqnarray}
In all one-loop expressions of radiative corrections, 
we adopt a (modified) minimal subtraction procedure
to remove poles in $\epsilon$
and keep only finite (real) parts of the above functions.

Some useful expressions for these functions in limiting cases
are (after minimal subtraction)
\begin{eqnarray}
&&B_0(0,m^2_1,m^2_2)\ =\ 1-\ln{m_1^2\over Q^2}+{m_2^2\over m_1^2-m_2^2}
\ln{m_2^2\over m_1^2}\ ,\label{B01}\\
&&B_0(m_1^2,m_2^2,0)\ =\ 2-\ln{m_1^2\over Q^2}
-\left(1-{m_2^2\over m_1^2}\right)\ln\left({1-{m_2^2\over m_1^2}}\right)
-{m_2^2\over m_1^2}\ln{m_2^2\over m_1^2}\ ,\label{B02}\\
&&\left.{d\over dp^2}B_0(p^2,m^2,m^2)\right|_{p^2=0}\ =\ {1\over6m^2}\ ,
\label{B03}\\
&&B_0(m^2,m^2,m^2)\ =\ -\ln{m^2\over Q^2} + 2 -{\pi\over\sqrt{3}}\ .\label{B04}
\end{eqnarray}

\subsection*{A.2 Two-loop scalar functions}

In this subsection we collect some useful 
formulae of zero-point two-loop scalar
functions. They have been studied extensively by several groups 
using two different methods: a differential equation method 
\cite{FJJ,CCLR} and an integral Mellin-Barnes transformation method \cite{DT};
their results all agree. Here we mainly follow Ref.~\cite{FJJ}.

The momentum integrals appearing in a two-loop 
effective potential calculation can be reduced to the following
two types of scalar functions [corresponding to the topologies of
two distinct zero-point two-loop irreducible Feynman diagrams (the 
figure-8 and sunset diagrams)]:
\begin{equation}
J(m_1^2,m_2^2)\ =\ -(\olf\mu^{4-d})^2
\int {d^dp~d^dq\over (2\pi)^{2d}}
{1\over [p^2 -m_1^2 + i\varepsilon][q^2 -m_2^2 + i\varepsilon]}\ ,
\end{equation}
and
\begin{equation}
I(m_1^2,m_2^2,m_3^2)\ =\ (\olf\mu^{4-d})^2
\int {d^dp~d^dq\over (2\pi)^{2d}}
{1\over [p^2 -m_1^2 + i\varepsilon][q^2 -m_2^2 + i\varepsilon]
[(p+q)^2-m_3^2+i\varepsilon]}
\ .
\end{equation}
The function $J$ is symmetric in $m_1, m_2$ and $I$ symmetric in
$m_1,m_2$ and $m_3$. 

The function $J$ can be reduced to the product of one-loop scalar
functions as 
\begin{equation}
J(m_1^2,m_2^2)\ =\ A_0(m_1^2) A_0(m_2^2)\ .
\end{equation}
The function $I$ satisfies the following first-order partial
differential equation \cite{CCLR}
\begin{eqnarray}
&&R^2~{\partial\over\partial m^2_3} I(m_1^2,m_2^2,m_3^2) \ =\ 
(d-3) (m_3^2-m_1^2-m_2^2) I(m_1^2,m_2^2,m_3^2) \nn\\
&+& (d-2)\biggl[{m_3^2-m_1^2+m_2^2\over 2m_3^2} J(m_1^2,m_3^2)
+ {m_3^2+m_1^2-m_2^2\over 2m_3^2} J(m_2^2,m_3^2)
- J(m_1^2,m_2^2)\biggr]\ ,\label{dIdm}
\end{eqnarray}
where 
\begin{equation}
R^2\ =\ m_1^4+m_2^4+m_3^4-2m_1^2 m_2^2 - 2 m_1^2 m_3^2 - 2 m_2^2 m_3^2\ .
\end{equation}
This differential equation can be used to solve for the $I$ function.
The initial value of this function can be evaluated from
(\ref{dIdm}) which reduces to a simple algebraic equation when 
$m_3=m_1+m_2$, {\it i.e.} $R=0$.

In our calculation, any Feynman diagram in the two-loop effective
potential is subtracted by all its possible one-loop
sub-diagrams; this is done by replacing 
the $I$ and $J$ functions as follows \cite{FJJ}:
\begin{eqnarray}
I(m_1^2,m_2^2,m_3^2) &\rightarrow & {\hat I}(m_1^2,m_2^2,m_3^2)
=I(m_1^2,m_2^2,m_3^2)
-{1\over{\epsilon}}\biggl[A_0(m_1^2)+A_0(m_2^2)+A_0(m_3^2)\biggr]\ ,\nn\\
J(m_1^2,m_2^2) &\rightarrow & {\hat J}(m_1^2,m_2^2)=J(m_1^2,m_2^2)
+{1\over{\epsilon}}\biggl[m_1^2A_0(m_2^2)+m_2^2A_0(m_1^2)\biggr]\ .
\label{Idef}
\end{eqnarray}
It is then straightforward to show
\begin{equation}
{\hat J}(m_1^2,m_2^2)\ =\ -{m_1^2 m_2^2\over{\epsilon}^2}
+m_1^2 m_2^2\biggl(1-\ln {m_1^2\over Q^2}\biggr)
\biggl(1-\ln{m_2^2\over Q^2}\biggr)\ ,\label{J}
\end{equation}
and with some effort
\begin{eqnarray}
{\hat I}(m_1^2,m_2^2,m_3^2) &=& {1\over2{\epsilon}^2}(m_1^2+m_2^2+m_3^2)
-{1\over2{\epsilon}}(m_1^2+m_2^2+m_3^2)\nn\\
&-&{1\over 2}\biggl[
(-m_1^2+m_2^2+m_3^2)
\ln{m^2_2\over Q^2}\ln{m^2_3\over Q^2}
+(m_1^2-m_2^2+m_3^2) \ln{m^2_1\over Q^2}\ln{m^2_3\over Q^2}
\nn\\
&+&(m_1^2+m_2^2-m_3^2) \ln{m^2_1\over Q^2}\ln{m^2_2\over Q^2}
-4\biggl(m^2_1\ln {m^2_1\over Q^2}+m^2_2\ln {m^2_2\over Q^2}
+m^2_3\ln {m^2_3\over
Q^2}\biggr)\nn\\
&+&\xi(m_1^2,m_2^2,m_3^2)+5(m^2_1+m^2_2+m^2_3)
\biggr]\ ,
\label{I}
\end{eqnarray}
where (for $R^2>0$) $\xi$ is given by \cite{DT} 
\begin{eqnarray}
\xi(m_1^2,m_2^2,m_3^2) 
&=& R\biggl[2\ln\biggl({m_3^2+m_1^2-m_2^2-R\over2m_3^2}\biggr)
\ln\biggl({m_3^2-m_1^2+m_2^2-R\over2m_3^2}\biggr)
-\ln{m_1^2\over m_3^2}\ln{m_2^2\over m_3^2}\nn\\
&-&2 Li_2\biggl({m_3^2+m_1^2-m_2^2-R\over2m_3^2}\biggr)
-2 Li_2\biggl({m_3^2-m_1^2+m_2^2-R\over2m_3^2}\biggr)
+{\pi^2\over3}\biggr]\ ,
\label{xi}
\end{eqnarray}
where $Li_2(x)$ is the dilogarithm function
\begin{equation}
Li_2(x)\ =\ -\int_0^1dy\ {\ln(1-x y)\over y}\ .
\end{equation}
In the region where $R^2<0$, (\ref{xi}) should be replaced by its
analytical continuation. Equivalent expressions for $\xi$ also appear
in \cite{FJJ} and \cite{CCLR}; we find that (\ref{xi})
is most convenient for series expansions.
We also define a function $L$ for future use
\be
L(m^2_1,m^2_2,m^2_3)\ =\ J(m^2_2, m^2_3) - J(m^2_1, m^2_2) - J(m^2_1, m^2_3)
-(m_1^2-m_2^2-m_3^2) I(m^2_1,m^2_2,m^2_3)\ .\label{L}
\ee

Performing a (modified) minimal subtraction (by removing the single and
double poles in $\epsilon$), it is the finite (real) parts of
(\ref{J}) and (\ref{I}) that we use in our 
two-loop effective potential expressions. 
We will also omit the carets 
of $\hat{I}$ and $\hat{J}$ to simplify the notation.

When computing the two-loop potential, some 
argument of the $I$ function, {\it e.g.}
the bottom-quark mass $m_b$, tree-level Higgs boson mass $m_{h^0}$, 
can be taken to be zero. The function $I$ is well-behaved
in these limiting cases:
\begin{eqnarray}
I(m_1^2,m_2^2,0) &=& 
-m_1^2\ln{m_1^2\over Q^2}\ln{m_2^2\over Q^2}
-(m_1^2-m_2^2)\ln{m^2_1-m_2^2\over Q^2}\ln {m_1^2\over m^2_2}
+{1\over2}(m_1^2-m_2^2)\ln^2{m_1^2\over Q^2}\nn\\
&+&2\biggl(m_1^2 \ln{m_1^2\over Q^2}+m_2^2 \ln{m_2^2\over Q^2}\biggr)
-{5\over2}(m_1^2+m_2^2)+(m_1^2-m_2^2)\biggl[
-{\pi^2\over6}+ Li_2 \left({m^2_2\over m^2_1}\right)\biggr]\ ,
\label{120}\nn\\
&&\\
I(m^2,0,0) &=& 
- m^2 \biggl({1\over2}\ln^2{m^2\over Q^2}	
-2\ln{m^2\over Q^2}+{5\over2}+{\pi^2\over6}\biggr)\ ,
\end{eqnarray}
where we have kept only the finite terms as explained before.
In (\ref{120}) we have implicitly assumed $m_1\geq m_2$. 
The symmetry of the above expression for $I(m_1^2,m_2^2,0)$
in $m_1$ and $m_2$ [which obviously follows from the definition
(\ref{Idef}) of $I$]
can be explicitly checked 
by using the identity 
\begin{equation}
Li_2(x)\ =\ -Li_2(x^{-1})-{1\over2}\ln^2(-x)-{\pi^2\over6}\ .
\end{equation}

Finally, we collect expansion formulae for the function $\xi$ which we
use in the derivation of an analytical approximation formula for the
two-loop Higgs boson mass corrections. The $\xi$ functions we find can be
reduced to one of the different types we list below using the relation
\begin{equation}
\xi(m_1^2,m_2^2,m_3^2)\ =\ m_1^2\xi(1,m_2^2/m_1^2,m_3^2/m_1^2)\ .
\end{equation}

(1) For $0\leq r\leq 1$ and $0\leq\epsilon\ll 1$:
\begin{eqnarray}
\xi(1,r,\epsilon) &=&(1-r)\left\{
\frac{\pi^2}{3}+\biggl[\ln\epsilon-2\ln(1-r)\biggr]\ln r-2 Li_2(r)
\right\}\nonumber\\
&-&\epsilon\left\{2-2\ln\epsilon+\ln r
+\frac{1+r}{1-r}\biggl[\frac{\pi^2}{3}
+\biggl(\ln\epsilon-2\ln(1-r)-1\biggr)\ln r-2Li_2(r)\biggr]
\right\}\nn\\
&+&\frac{\epsilon^2}
{(1-r)^3}\left\{\left(\frac{3}{2}-\ln\epsilon\right)(1-r^2)
-\frac{2\pi^2}{3}r-
\biggl[2\ln\epsilon+r-4\ln(1-r)\biggr]r\ln r\right.\nn\\
&&\qquad\qquad \left.
+4r Li_2(r)\frac{}{}\right\}+{\cal O}(\epsilon^3)\ .
\label{xiapp1}
\eear
If $r>1$, one uses $\xi(1,r,\epsilon)=r\xi(1,1/r,\epsilon/r)$ and the
above expression.

Two particular cases of the previous expansion are:

(1a) For $0\leq\epo,\ept\ll 1$:
\begin{eqnarray}
\xi(1,\epo,\ept) &=& {\pi^2\over3}+\ln\epo\ln\ept
-2\biggl(-1+{\pi^2\over3}+\ln\epo\ln\ept\biggr)\epo\ept \\
&+&\biggl[\biggl(-2-{\pi^2\over3}+2\ln\epo-\ln\epo\ln\ept\biggr)\epo+
\biggl({3\over2}-\ln\epo\biggr)\epo^2+(\epo\leftrightarrow\ept)\biggr]
+{\cal O}(\epo^m\ept^n)\ ,\nn
\label{xiapp1a} 
\end{eqnarray}
with $m+n=3$, and

(1b) For $0\leq |\epo|, \ept\ll 1$:   
\begin{eqnarray}
\xi(1,1+\epo,\ept) &=& -2(4+\epo-2\ln\ept)\ept
+\biggl({8\over9}-{1\over3}\ln\ept\biggr)
\ept^2 \nn\\
&+&\biggl[2-\ln\ept+\biggl({7\over18}
+{1\over6}\ln\ept\biggr)\ept\biggr]\epo^2\nn\\
&+&\biggl(-\frac{1}{2}+{1\over2}\ln\ept\biggr)\epo^3
+\biggl({2\over9}-{1\over3}\ln\ept\biggr)\epo^4
+{\cal O}(\epo^m\ept^n)\ ,
\label{xiapp1b}
\end{eqnarray}
with $m+2 n\geq 5$.

Finally we also give

(2) For $|\epo|, |\ept|\ll 1$:
\begin{eqnarray}
\xi(1,1+\epo,1+\ept)&=&
36 K + (8K-1)\epo\ept +\biggl({5\over36}-{8\over3}K\biggr)\epo^2\ept^2\nn\\
&+& \biggl\{ 12 K\epo + (1-8 K)\epo^2+\biggl({8\over3}K-{2\over9}\biggr)\epo^3
+\biggl({1\over108}-{16\over9}K\biggr)\epo^4\nn\\
&+&\biggl[-{\epo^2\over6}
+\biggl({11\over54}+{8\over9}K\biggr)\epo^3\biggr]\ept
+(\epo\leftrightarrow\ept)\biggr\}+{\cal O}(\epo^m\ept^n)\ ,
\label{xiapp2}
\end{eqnarray}
with $m+n=5$. In this expansion the constant number $K$ is given by 
\begin{equation}
K\ =\ -{1\over\sqrt{3}}\int_0^{\pi/6}dx\ \ln(2\cos x)\simeq -0.1953256\ .
\end{equation}

\section*{Appendix B: MSSM in the leading approximation} 
\setcounter{equation}{0}
\renewcommand{\theequation}{B.\arabic{equation}}

The general structure of the MSSM is quite complicated, with many
different fields and field mixings. This makes the computation of the
complete potential prohibitive at two-loops.
However, it is a good aproximation to keep only those 
terms of the MSSM Lagrangian which depend on the $SU(3)$ gauge coupling 
$g_3$ and the top Yukawa $h_t$
(and neglect the electroweak gauge couplings
$g_1,g_2$ and the rest of the Yukawa couplings).
We call this the leading approximation  and it greatly simplifies our 
two-loop effective potential calculation. In this Appendix,
we summarize the necessary Feynman rules for computing the
two-loop potential in this leading approximation
and also some MSSM renormalization group equations, useful to check the
scale invariance of the potential.

\subsection*{B.1 Masses and Feynman rules}

The Higgs sector scalar potential in the leading approximation is
\begin{equation}
V_{\rm Higgs}\ =\ (m^2_{H_1}+\mu^2)|H_1|^2 + (m^2_{H_2}+\mu^2)|H_2|^2 
+B_\mu (H_1 H_2 +{\rm H.c.})\ , 
\label{mssm}
\end{equation}
where $m_{H_1}, m_{H_2}$ and $B_\mu$ 
[with dimensions of (mass)$^2$]
are the soft-breaking Higgs mass parameters,
and $\mu$ the supersymmetric Higgs-boson mass parameter.
Although we do not write the quartic Higgs couplings, which depend on the
electroweak gauge coupling constants, they are responsible for the
tree-level mass of the lightest Higgs boson, which we of course include in
our calculations.

The $SU(2)$ doublet Higgs fields $H_1$ and $H_2$ can be written as follows:
\begin{equation}
H_1=\left[\begin{array}{c}
{(h_1 + i a_1)/{\sqrt 2}}\\[2mm]
 h_1^-
\end{array}\right]\ , \qquad
H_2=\left[\begin{array}{c}
h_2^+\\[2mm]
{(h_2 + i a_2)/{\sqrt 2}}
\end{array}\right]\ .
\label{h1h2}
\end{equation}
In our approximation, the mass-squared matrices for ${\cal CP}$-even and
odd Higgs fields are
\begin{equation}
{\cal M}^2_{\pm}\ =\ 
\left(
\begin{array}{cc}
m^2_{H_1}+\mu^2 & \pm B_\mu\\
\pm B_\mu & m^2_{H_2}+\mu^2
\end{array}\right)\ ,\label{msqeven}
\end{equation}
where the positive (negative) sign applies to the ${\cal CP}$-even
(odd) fields respectively.
The charged Higgs fields have the same mass-squared matrix 
${\cal M}^2_{-}$ as the ${\cal CP}$-odd Higgses.

The ${\cal CP}$-even interaction eigenstates $h_1,~h_2$ are rotated by the
angle
$\alpha$ into the mass eigenstates $H^0$ and $h^0$. 
Similarly, the ${\cal CP}$-odd states $a_1,~a_2$ (charged
states $h^+_1,~h^+_2$) are rotated into mass
eigenstates $G^0$ and $A^0$ ($G^+$ and $H^+$) 
by the angle $\beta$. This angle $\beta$ is conventionally defined
in terms of the ${\cal CP}$-even 
Higgs field VEVs, $\langle h_{1,2}\rangle =v_{1,2}$,
by $\tan\beta=v_2/v_1$. The fact that $\beta$ diagonalizes 
${\cal M}^2_{-}$ is obvious when the minimization conditions
of the potential (\ref{mssm}), $m_{H_1}^2+\mu^2 = -B_\mu\tan\beta$
and $m_{H_2}^2+\mu^2 = -B_\mu\cot\beta$,
are imposed and the soft parameters
in the matrix are replaced by $\tan\beta$ and
$m^2_{A^0} = - B_\mu(\tan\beta+\cot\beta)$. Since 
we have neglected all $g_1,~g_2$ related terms in (\ref{mssm}),
in our approximation (we use shorthand notations $c_\beta=\cos\beta$, 
$s_\beta=\sin\beta$, {\it etc.})
\begin{equation}
c_\alpha\ =\ s_\beta,~~{\rm and}~s_\alpha\ =\ -c_\beta\ .
\label{alphabeta}
\end{equation}
This approximation is excellent when $M_{A^0}\gg M_Z$ but would fail
for $M_{A^0}\sim M_Z$. The effect is numerically relevant for the tree
level masses and we take it into account, but it may be consistently
neglected in the two-loop corrections.

The (field-dependent) top and bottom squark mass-squared matrices 
(neglecting the $D$-terms) are\footnote{In this revised version,
we have also included bottom Yukawa terms in the Feynman rules, 
they will be used in the  expanded two-loop effective potential expression
(\ref{2lat}).}
\begin{equation}
{\cal M}^2_{\tilt}\ =\ 
\left[\begin{array}{cc}
\MQ^2+{1\over2}h_t^2h^2_2
	& {1\over\sqrt{2}}h_t(A_th_2+\mu h_1)\\[2mm]
{1\over\sqrt{2}}h_t(A_th_2+\mu h_1) 
	& \MU^2+{1\over2}h_t^2h^2_2
\end{array}\right]\ ,\label{squarkm}
\end{equation}
\begin{equation}
{\cal M}^2_{\tilb}\ =\ 
\left[\begin{array}{cc}
\MQ^2+{1\over2}h_b^2h^2_1
	& {1\over\sqrt{2}}h_b(A_b h_1+\mu h_2)\\[2mm]
{1\over\sqrt{2}}h_b(A_bh_1+\mu h_2) 
	& \MD^2+{1\over2}h_b^2h^2_1
\end{array}\right]\ ,\label{sbquarkm}
\end{equation}
where $\MQ$, $\MU$ ($\MD$) are soft-breaking mass parameters 
of the left- and right-handed top(bottom)-squarks
$\widetilde Q$ and $\widetilde U$ 
($\widetilde D$); $A_t$  and $A_b$ are the usual trilinear 
soft-breaking parameters. We denote the mass eigenvalues of the matrix
(\ref{squarkm}) by $m_{\tilt_1}$, $m_{\tilt_2}$ 
and the mixing angle by $\theta_{\tilt}$, and the corresponding quantities for
the matrix (\ref{sbquarkm}) by $m_{\tilb_1}$, $m_{\tilb_2}$ and
$\theta_{\tilb}$. 

The Feynman rules for Higgs/Goldstone-boson-squark trilinear coupling are
simply $-i\lambda$, with $\lambda$ as listed below:
\begin{eqnarray}
\lambda_{H^+\tilt_1\tilb_1}&=& 
-h_t \cbe \biggl[ (c_t m_t + s_t Y_t)c_b+m_b s_b s_t\biggl]
-h_b \sbe \biggl[ (c_b m_b + s_b Y_b)c_t+m_t s_t s_b\biggl]
\ ,\nn\\
\lambda_{H^+\tilt_1\tilb_2}&=& \;\;\;
h_t \cbe \biggl[ (c_t m_t + s_t Y_t)s_b-m_b c_b s_t\biggl]
+h_b \sbe \biggl[ (s_b m_b - c_b Y_b)c_t-m_t s_t c_b\biggl]
\ ,\nn\\
\lambda_{H^+\tilt_2\tilb_1}&=& \;\;\;
h_t \cbe \biggl[ (s_t m_t - c_t Y_t)c_b-m_b s_b c_t\biggl]
+h_b \sbe \biggl[ (c_b m_b + s_b Y_b)s_t-m_t c_t s_b\biggl]
\ ,\nn\\
\lambda_{H^+\tilt_2\tilb_2}&=&
-h_t \cbe \biggl[ (s_t m_t - c_t Y_t)s_b+m_b c_b c_t\biggl] 
-h_b \sbe \biggl[ (s_b m_b - c_b Y_b)s_t+m_t c_t c_b\biggl]
\label{hstsb}\ ,
\end{eqnarray}
\begin{eqnarray}
\lambda_{G^+\tilt_1\tilb_1}&=& 
-h_t \sbe (c_t m_t + s_t X_t)c_b
+h_b \cbe (c_b m_b + s_b X_b)c_t
\ ,\nn\\
\lambda_{G^+\tilt_1\tilb_2}&=& \;
h_t \sbe (c_t m_t + s_t X_t)s_b
+h_b \cbe (-s_b m_b + c_b X_b)c_t
\ ,\nn\\
\lambda_{G^+\tilt_2\tilb_1}&=& \;
h_t \sbe (s_t m_t - c_t X_t)c_b
-h_b \cbe (c_b m_b + s_b X_b)s_t
\ ,\nn\\
\lambda_{G^+\tilt_2\tilb_2}&=&
h_t \sbe  (-s_t m_t + c_t X_t)s_b
+h_b \cbe (s_b m_b - c_b X_b)s_t
\label{Gstsb}\ ,
\end{eqnarray}
and
\begin{eqnarray}
\lambda_{H^0\tilt_1\tilt_1}= \sqrt{2}h_t(m_t + s_tc_tY_t^\alpha)\sa\ ,
&&
\lambda_{H^0\tilt_2\tilt_2}= \sqrt{2}h_t
(m_t - s_t c_t Y_t^\alpha)\sa\ ,\nn\\
\lambda_{h^0\tilt_1\tilt_1}= \sqrt{2}h_t
(m_t + s_tc_t X_t^\alpha)\ca\ ,
&&
\lambda_{h^0\tilt_2\tilt_2}= \sqrt{2}h_t
(m_t - s_t c_t X_t^\alpha)\ca\ ,\nn\\
\lambda_{H^0\tilt_1\tilt_2}=\frac{1}{\sqrt{2}}h_t c_{2t}Y_t^\alpha\sa\ ,
&&
\lambda_{h^0\tilt_1\tilt_2}=\frac{1}{\sqrt{2}}h_t c_{2t}X_t^\alpha\ca\ ,
\nn\\
\lambda_{A^0\tilt_1\tilt_2}=-\lambda_{A^0\tilt_2\tilt_1}
=\frac{1}{\sqrt{2}}h_tY_t\cbe\ ,
&&
\lambda_{G^0\tilt_1\tilt_2}=-\lambda_{G^0\tilt_2\tilt_1}=
\frac{1}{\sqrt{2}}h_t X_t\sbe\ ,\label{hstst}
\end{eqnarray}
where $c_t=\cos\theta_{\tilt}$, $s_t=\sin\theta_{\tilt}$,
$c_{2t}=\cos2\theta_{\tilt}$ (with similar expressions for 
$\theta_{\tilb}$ functions) and
\begin{equation}
X_t\ =\ A_t+\mu\cot\beta,\qquad Y_t\ =\ A_t - \mu\tan\beta\ ,
\label{Xtdef}
\end{equation}
\begin{equation}
X_b\ =\ A_b+\mu\tan\beta,\qquad Y_b\ =\ A_b - \mu\cot\beta\ .
\label{Xbdef}
\end{equation}
In addition, we find convenient to define the $\alpha$-dependent quantities
\begin{equation}
X_t^\alpha\ =\ A_t-\mu\tan\alpha,\qquad Y_t^\alpha\ =\ A_t + \mu\cot\alpha\ ,
\label{Xtadef}
\end{equation}
\begin{equation}
X_b^\alpha\ =\ A_b-\mu\cot\alpha,\qquad Y_b^\alpha\ =\ A_b + \mu\tan\alpha\ ,
\label{Xbadef}
\end{equation}
which tend to the corresponding quantities without the
$\alpha$ label ($X_t^\alpha\rightarrow X_t$, etc) in the limit $m_A\gg M_Z$. 

Couplings similar to the above ones but for bottom squarks can be obtained
directly from (\ref{hstsb}), (\ref{Gstsb}) and (\ref{hstst}): simply make
everywhere the replacements $\{ h_t\lra h_b,\, m_t\lra m_b,\,
\theta_{\tilt}\lra \theta_{\tilb},\, X_t^{(\alpha)}\lra X_b^{(\alpha)},\,
Y_t^{(\alpha)}\lra Y_b^{(\alpha)}\}$ and $\{\ca \lra\sa,\, \cbe\lra -\sbe
\}$ for the couplings to $\{H^+, A^0, H^0\}$ or $\{\ca \lra-\sa,\,
\cbe\lra \sbe \}$ for the couplings to $\{G^+, G^0, h^0\}$.

The couplings of squarks to neutralinos and charginos
are very simple in the
leading approximation, since the gaugino-Higgsino mixing can be
neglected and the only interactions are Higgsino-squark
interactions. 
The Feynman rules for the ${\tilde h_i^0}t{\tilde t}_j$
couplings can be written as $-i(a{\cal P}_L+b{\cal P}_R)$
and that of ${\tilde h^+}t{\tilde b}_L$ 
as $i{\cal C}^{-1}(a{\cal P}_L+b{\cal P}_R)$
(${\cal P}_{L,R}$ are chiral projectors 
and ${\cal C}$ the charge-conjugation matrix), with
\begin{eqnarray}
&&a_{{\tilde h_1^0}t{\tilde t}_1}=-ia_{{\tilde h_2^0}t{\tilde t}_1} 
 = b_{{\tilde h_1^0}t{\tilde t}_2} = ib_{{\tilde h_2^0}t{\tilde t}_2} 
 = {h_t\over\sqrt{2}}c_t\ ,\nn\\
&&-a_{{\tilde h_1^0}t{\tilde t}_2} = ia_{{\tilde h_2^0}t{\tilde t}_2} 
 = b_{{\tilde h_1^0}t{\tilde t}_1} = ib_{{\tilde h_2^0}t{\tilde t}_1} 
 = {h_t\over\sqrt{2}}s_t \ ,\nn\\
&&a_{{\tilde h^+}t{\tilde b}_L} = -h_t\ ,
\qquad
a_{{\tilde h^+}b{\tilde t}_1} = -h_ts_t\ ,
\qquad
a_{{\tilde h^+}b{\tilde t}_2} = -h_tc_t\ ,
\end{eqnarray}
when $\mu>0$; for $\mu<0$, we only need to interchange 
$a_{{\tilde h_1^0}t{\tilde t}_i}$ and $a_{{\tilde h_2^0}t{\tilde t}_i}$,
as well as
$b_{{\tilde h_1^0}t{\tilde t}_i}$ and $b_{{\tilde h_2^0}t{\tilde t}_i}$.

Other Feynman rules of ${\cal O}(g_3)$ and ${\cal O}(h_t)$ vertices
are exactly the same as in the general MSSM and we do not present them
explicitly.

\subsection*{B.2 Renormalization group equations}

The MSSM RGEs \cite{martin} that we will use 
to check the invariance of the potential to two-loop order
under renormalization scale transformations are the following. First, we
need the two-loop RGEs
for those parameters entering in the tree-level potential (\ref{mssm})
\begin{eqnarray}
{\partial m_{H_2}^2\over\partial\ln Q^2} &=&~
{3h_t^2\over16\pi^2}{\cal M}^2_t +{16g_3^2h_t^2\over(16\pi^2)^2}
({\cal M}^2_t+2M_3^2-2M_3A_t)-{18h^4_t\over(\olf)^2}({\cal M}^2_t+A_t^2)
\ ,\label{mH2}\\
{\partial\ln\mu\over\partial\ln Q^2} &=& 
- {\partial\ln h_2\over\partial\ln Q^2}\ =\
{3h_t^2\over32\pi^2}+{8g_3^2h_t^2\over(16\pi^2)^2}-\frac{9}{2}
{h^4_t\over(\olf)^2}
\ ,\label{mu}\\
{\partial B_\mu\over\partial\ln Q^2} &=&
{3h_t^2\over16\pi^2}\left({B_\mu\over2}+A_t\mu\right)
+{16g_3^2h_t^2\over(16\pi^2)^2}
\left({B_\mu\over2}+A_t\mu-M_3\mu\right)\nn\\
&-&{9h^4_t\over(\olf)^2}\biggl({B_\mu\over2}+2 A_t\mu\biggr)\ ,\label{bmu}
\end{eqnarray}
where ${\cal M}^2_t = m^2_{H_2}+\MQ^2+\MU^2+A_t^2$. Then we need 
one-loop RGEs
for those masses entering in the one-loop potential
\begin{eqnarray}
\olf~{\partial m^2_t\over\partial\ln Q^2} &=& 
\biggl(-{16g_3^2\over3}+3h_t^2\biggr) m_t^2\ ,\label{mt}\\
\olf~{\partial m^2_{\tilt_1}\over\partial\ln Q^2} &=& 
-{16g_3^2\over3}\biggl[m_t^2+M_3^2- s_{2t} m_t\biggl(M_3 - {\xt\over2}\biggr)
\biggr]\nn\\
&+&h_t^2\biggl[ 3m_t^2+(1+s^2_t){\cal M}^2_t 
+ 3 s_{2t} m_t\biggl(A_t+{3\xt\over2}
\biggr)\biggr]\ ,\label{mst1}\\
\olf~{\partial m^2_{\tilt_2}\over\partial\ln Q^2} &=& 
-{16g_3^2\over3}\biggl[m_t^2+M_3^2 + s_{2t} m_t\biggl(M_3 - {\xt\over2}\biggr)
\biggr]\nn\\
&+&h_t^2\biggl[ 3m_t^2+(1+c_t^2){\cal M}^2_t 
- 3 s_{2t} m_t\biggl(A_t+{3\xt\over2}
\biggr)\biggr]\ ,\label{mst2}\\
\olf~{\partial m^2_{H^0_n}\over\partial\ln Q^2}
&=& 3h_t^2\biggl[\mu^2 + D_n{\cal M}^2_t + E_n \biggl( {B_\mu\over2}
+A_t\mu\biggr)\biggr]\ ,\label{H0}\\
\olf~{\partial m^2_{H^+_n}\over\partial\ln Q^2}
&=& 3h_t^2\biggl[\mu^2 + D_{n+2}{\cal M}^2_t + E_{n+2} \biggl( {B_\mu\over2}
+A_t\mu\biggr)\biggr]\ ,\label{H+}
\end{eqnarray}
where $D_n=\sa^2, \ca^2, \sbe^2, \cbe^2$ and
$E_n=s_{2\alpha},-s_{2\alpha}, -s_{2\beta}, s_{2\beta}$ 
for $n=1,2,3,4$. [Here we use the $\alpha$ angle to keep track the
$H^0$ and $h^0$ contributions; it  can
be replaced by the $\beta$ angle as in (\ref{alphabeta})
in the leading approximation.]
The ordering of the Higgs/Goldstone bosons are
$H_n^0=H^0, h^0, G^0$ and $A^0$ for $n=1,2,3,4$ and 
$H_n^+=G^+, H^+$ for $n=1,2$. 
Eqs.~(\ref{mst1}-\ref{H+}) seem unfamiliar, but they
follow directly from (\ref{msqeven}), (\ref{squarkm})
and the one-loop MSSM RGEs of soft parameters entering those equations. 

Using (\ref{mst1}) and (\ref{mst2}),
we find one-loop RGEs for $X_t$ and $m^2_{\tilt}$,
the (arithmetic) average of the (squared) top squark masses. They are
\begin{eqnarray}
\olf~{\partial X_t\over\partial\ln Q^2} &=& 
{16\over3}g_3^2M_3 + 3 h_t^2(A_t+X_t)\ .\label{Xt}\\
\olf~{\partial m^2_{\tilt}\over\partial\ln Q^2} &=& 
-{16\over3}g_3^2(m_t^2+M_3^2)
+h_t^2\biggl(3m_t^2+{3\over2}{\cal M}^2_t\biggr)\ ,\label{mst}
\end{eqnarray}
these two equations are used in Sec.~3 for the RG discussion of the
formula for the Higgs boson mass $M_{h^0}$. Eq.~(\ref{Xt}) can also be
derived from
(\ref{Xtdef}) and one-loop RGEs of $A_t$, $\mu$ and $\tan\beta$.

\section*{Appendix C: One-loop self-energies} 
\setcounter{equation}{0}
\renewcommand{\theequation}{C.\arabic{equation}}

In this appendix, we collect formulae for those MSSM one-loop
self-energies which are necessary for our analysis. We present these
self-energies in the leading approximation of keeping only $h_t$ and
$g_3$-dependent terms, as explained in Appendix B; their full form can be
found in Ref.~\cite{PBMZ}, which we follow for notation. (See also
\cite{Donini} for top quark/squark self-energies.)

\noindent{\it --Top quark}:
\begin{eqnarray}
\olf~\Sigma_t(p^2) &=& {4g^2_3\over3}\Biggl\{
m_t\biggl[B_1(p^2,m^2_{\tilde g},m^2_{\tilt_1})
+B_1(p^2,m^2_{\tilde g},m^2_{\tilt_2})\biggr]
-m_t\biggl(5-3\ln{m^2_t\over Q^2}\biggr)\nn\\
&-&s_{2t}~m_{\tilde g}\biggl[B_0(p^2,m^2_{\tilde g},m^2_{\tilt_1})-
B_0(p^2,m^2_{\tilde g},m^2_{\tilt_2})\biggr]\Biggr\}\nn\\
&+& {h_t^2\over2}m_t \Biggl\{ c^2_\beta \biggl[2 B_1(p^2,m_t^2,m^2_{A^0})
+B_1(p^2,m^2_b,m^2_{A^0})\biggr]\nn\\
&+&s^2_\beta \biggl[2 B_1(p^2,m_t^2,m^2_Z)
+B_1(p^2,m^2_b,m^2_Z)\biggr]\nn\\
&+&B_1(p^2,\mu^2,m^2_{\tilt_1})
+B_1(p^2,\mu^2,m^2_{\tilt_2})
+B_1(p^2,\mu^2,m^2_{\tilb_L})\Biggr\}\ ,
\label{pit}
\end{eqnarray}
where we have assumed all heavy Higgs bosons have mass $m_{A^0}$
much larger than the masses of the light Higgs and $W$-boson,
taken to be $\sim m_Z$.

From (\ref{pit}) we find the running top-quark mass at the scale
$Q$ (under the simplified assumptions of a common heavy SUSY scale $M_S$
while the $\mu$ parameter is left free, see Sec. 2)
\begin{eqnarray}
m^2_t(Q) &=& M_t^2\Biggl\{1-{g_3^2\over 6\pi^2}\biggl[
5-3\lt+\ls-\hxt\biggr]\label{mtQ}\\
&+&{3h_t^2\over 32\pi^2}
\biggl[
(1+c^2_\beta)\biggl({1\over2}-\ls\biggr)
+s^2_\beta\biggl({8\over3}-\lt\biggr)
-\frac{\hmu^2}{1-\hmu^2}\biggl(
1+\frac{\hmu^2}{1-\hmu^2}\ln{\hmu^2}
\biggl)
\biggr]\Biggr\}\ .\nn
\end{eqnarray}
In this equation we have neglected the external momentum 
and used (\ref{B1}) and (\ref{B01}). We have used
the reduced parameters $\hat{X}_t\equiv X_t/M_S$,
$\hat{\mu}\equiv\mu/M_S$ and $M_t$ is the top quark
pole mass (we use capital letters to denote on-shell mass
parameters). 

\noindent{\it --Top squarks}:
\begin{eqnarray}
\olf~\Pi_{\tilt_L\tilt_L}(p^2) &=& {8g^2_3\over3}\Biggl\{
G(p^2,m^2_{\tilde g},m^2_t)+c^2_t\biggl[A_0(m^2_{\tilt_1})
-(p^2+m^2_{\tilt_1})B_0(p^2,m^2_{\tilt_1},0)\biggr]\nn\\
&+&s^2_t\biggl[A_0(m^2_{\tilt_2})
-(p^2+m^2_{\tilt_2})B_0(p^2,m^2_{\tilt_2},0)\biggr]\Biggr\}\nn\\
&+&h^2_t\biggl[s^2_t A_0(m^2_{\tilt_1})+c^2_t A_0(m^2_{\tilt_2})
+{1\over2}\sum_{n=1}^4 D_n A_0(m^2_{H_n^0})+G(p^2,\mu^2,m_t^2)\biggr]\nn\\
&+&\sum_{n=1}^4\sum_{i=1}^2\lambda^2_{H^0_n\tilt_L\tilt_i}
B_0(p^2,m^2_{H^0_n},m^2_{\tilt_i})
+\sum_{n=1}^2\lambda^2_{H^+_n\tilt_L\tilb_L}
B_0(p^2,m^2_{H^+_n},m^2_{\tilb_L})\ ,\label{piLL}
\end{eqnarray}
\begin{eqnarray}
\olf~\Pi_{\tilt_R\tilt_R}(p^2) &=&
{8g^2_3\over3}\Biggl\{
G(p^2,m^2_{\tilde g},m^2_t)+s^2_t\biggl[A_0(m^2_{\tilt_1})
-(p^2+m^2_{\tilt_1})B_0(p^2,m^2_{\tilt_1},0)\biggr]\nn\\
&+&c^2_t\biggl[A_0(m^2_{\tilt_2})
-(p^2+m^2_{\tilt_2})B_0(p^2,m^2_{\tilt_2},0)\biggr]\Biggr\}\nn\\
&+&h^2_t\biggl[c^2_t A_0(m^2_{\tilt_1})+s^2_t A_0(m^2_{\tilt_2})
+A_0(m^2_{\tilb_L})\nn\\
&+&{1\over2}\sum_{n=1}^4 D_n A_0(m^2_{H_n^0})
+\sum_{n=1}^2 D_{n+2} A_0(m^2_{H_n^+})
+G(p^2,\mu^2,m_t^2)+G(p^2,\mu^2,m^2_b)\biggr]\nn\\
&+&\sum_{n=1}^4\sum_{i=1}^2\lambda^2_{H^0_n\tilt_R\tilt_i}
B_0(p^2,m^2_{H^0_n},m^2_{\tilt_i})
+\sum_{n=1}^2\lambda^2_{H^+_n\tilt_R\tilb_L}
B_0(p^2,m^2_{H^+_n},m^2_{\tilb_L})\ ,\label{piRR}
\end{eqnarray}
\begin{eqnarray}
\olf~\Pi_{\tilt_L\tilt_R}(p^2) &=&
{4 g^2_3\over3}\biggl[
-s_{2t}(p^2+m_{\tilt_1}^2)B_0(p^2,m^2_{\tilt_1},0)
+s_{2t}(p^2+m_{\tilt_2}^2)B_0(p^2,m^2_{\tilt_2},0)\nn\\
&+&4m_{\tilde g}m_t B_0(p^2,m^2_{\tilde g}, m^2_t)\biggr]
+{3\over2}h^2_t s_{2t}\biggl[A_0(m^2_{\tilt_1})-A_0(m^2_{\tilt_2})\biggr]\nn\\
&+&\sum_{n=1}^4\sum_{i=1}^2\lambda_{H^0_n\tilt_L\tilt_i}
\lambda_{H^0_n\tilt_R\tilt_i}
B_0(p^2,m^2_{H^0_n},m^2_{\tilt_i})\nn\\
&+&\sum_{n=1}^2\lambda_{H^+_n\tilt_L\tilb_L}\lambda_{H^+_n\tilt_R\tilb_L}
B_0(p^2,m^2_{H^+_n},m^2_{\tilb_L})\ ,\label{piLR}
\end{eqnarray}
where 
$\lambda_{H^0\tilt_L\tilt_1}=c_t\lambda_{H^0\tilt_1\tilt_1}
-s_t\lambda_{H^0\tilt_2\tilt_1}$, 
$\lambda_{H^0\tilt_R\tilt_1}=s_t\lambda_{H^0\tilt_1\tilt_1}
+c_t\lambda_{H^0\tilt_2\tilt_1}$,
{\it etc.}. The symbols $D_n$ are defined after (\ref{Xtdef}).

From (\ref{piLL}-\ref{piLR}) we derive relations between
running and on-shell top-squark masses and mixing parameters 
using the following one-loop relationships (for $c_t^2=s_t^2=1/2$):
\begin{eqnarray}
M_{\tilde{t}_1}^2&=&\MQ^2+m_t^2+m_tX_t
-\frac{1}{2}{\rm Re}\left[\Pi_{\tilt_L\tilt_L}(M^2_{\tilt_1})
+\Pi_{\tilt_R\tilt_R}(M^2_{\tilt_1})\right]-
{\rm Re}~\Pi_{\tilt_L\tilt_R}(M^2_{\tilt_1})\ ,\nn\\
M_{\tilde{t}_2}^2&=&\MQ^2+m_t^2-m_tX_t
-\frac{1}{2}{\rm Re}\left[\Pi_{\tilt_L\tilt_L}(M^2_{\tilt_2})
+\Pi_{\tilt_R\tilt_R}(M^2_{\tilt_2})\right]+
{\rm Re}~\Pi_{\tilt_L\tilt_R}(M^2_{\tilt_2})\ ,
\end{eqnarray}
we obtain (assuming again a common heavy SUSY scale $M_S$ and leaving
free the $\mu$-parameter):
\begin{eqnarray}
m^2_{\tilt}(Q) &=& 
M_{\tilt}^{2}\Biggl\{1-{g_3^2\over 3\pi^2}\left(2-\ls\right)
+{3h_t^2\over32\pi^2}
\biggl[(\hxt^2 s^2_\beta+\hyt^2 c^2_\beta)
\left(2-\ls\right)\nn\\
&+&c^2_\beta\left(1-{\pi\over\sqrt{3}}\hyt^2-\ls\right)\nn\\
&+&\hmu^4\ln{\hmu^2}+(1-\hmu^2)\left(3-2\ls\right)
-(1-\hmu^2)^2\ln(1-\hmu^2)
\biggr]\Biggr\}\ ,\\
m_t X_t (Q) &=& M_t X^{\rm OS}_t +{g_3^2\over
12\pi^2}m_tM_S\biggl[4(2-\ls)+2\hxt \ls\biggr]\nn\\
&+& {3 h_t^2\over\olf} m_t 
\biggl\{(X_t s^2_\beta+Y_t c^2_\beta)\left(2-\ls\right)
- {\pi\over\sqrt{3}}Y_t c_\beta^2+X_t\left(1-{3\over2}\ls\right) \nn\\
&-&\frac{1}{2}
\biggl[1-\hmu^2+\hmu^4\ln{\hmu^2}+(1-\hmu^4)\ln(1-\hmu^2) 
\biggr]
X_t \nn\\
&+& \biggl(-{1\over2} + {\pi\over3\sqrt{3}}\biggr)\cbe^2\hyt^2 X_t 
- {1\over2}\sbe^2\hxt^2 X_t \ln\left({m_t X_t\over M_S^2}\right)
- {1\over3}\sbe^2\hxt^2 X_t\ln2 \Biggr\}\ ,
\end{eqnarray}
where we have used (\ref{B02}), (\ref{B04}) and the definition 
$\hat{Y}_t\equiv (A_t-\mu\tan\beta)/M_S$.

\noindent{\it --$W$ boson}:
\begin{eqnarray}
\olf~\Pi^T_{WW}(p^2) &=& 3 g^2\Biggl\{ 2 B_{22}(p^2,m^2_t,m^2_b)
+{1\over2} G(p^2,m_t^2,m_b^2)-2 c^2_t \biggl[B_{22}(p^2,m_{\tilt_1}^2,
m^2_{\tilb_L})-{1\over4}A_0(m_{\tilt_1}^2)\biggr]\nn\\
&-&2 s^2_t \biggl[B_{22}(p^2,m_{\tilt_2}^2,
m^2_{\tilb_L})-{1\over4}A_0(m_{\tilt_2}^2)\biggr]
+{1\over2}A_0(m^2_{\tilb_L})\Biggr\}\ .\label{piWW}
\end{eqnarray}
This gives (under the assumption of the  simplified SUSY spectrum of
Sec.~2, described already for previous self-energies)
\begin{equation}
v^2(Q) = {4\over g^2} [M_W^2+{\rm Re}~\Pi_{WW}^T(M_W^2)]
= {4 M^2_W\over g^2}
\biggl[1-{h_t^2s_\beta^2\over32\pi^2}\left(-6\lt+3+\hxt^2\right)\biggr]\ ,
\label{v2Q}
\end{equation}
where we have neglected the external momentum in (\ref{piWW}) 
and used (\ref{B1}-\ref{B01}).

\noindent{\it --Higgs boson}:
We need only the difference 
\begin{eqnarray}
\olf~\biggl[\Pi_{hh}(m^2_{h^0})-\Pi_{hh}(0)\biggl]
&=& 3 h_t^2 m^2_{h^0} s^2_\beta\biggl[B_0(0,m_t^2,m_t^2)
-4 m^2_t \left. {d\over dp^2}B_0(p^2,m^2_t,m^2_t)\right|_{p^2=0}\biggr]\nn\\
&+& 3m^2_{h^0}\sum_{i,j}\lambda^2_{h^0\tilt_i\tilt_j}
\left. {d\over dp^2}B_0(p^2,m^2_{\tilt_i},m^2_{\tilt_j})\right|_{p^2=0}\ ,
\end{eqnarray}
where $\lambda_{h\tilt_i\tilt_j}$ are defined in (\ref{hstst}).
Using (\ref{B01}) and (\ref{B02}) we get (\ref{dpi}).

\section*{Appendix D: MSSM effective potential to the two-loop order} 
\setcounter{equation}{0}
\renewcommand{\theequation}{D.\arabic{equation}}

In this Appendix we present the MSSM effective potential for the (real)
neutral components of the Higgs fields up
to the two-loop level in the leading approximation (which neglects all
dimensionless couplings except $h_t$ and $g_3$). 
We first write the potential as
\begin{equation}
V(h_1,h_2)\ =\ V_{\rm vac}+V_0(h_1,h_2)+V_1(h_1,h_2)+V_2(h_1,h_2)\ ,
\end{equation}
where $V_{\rm vac}$ is a field-independent 
vacuum energy term\footnote{This term is a function of the soft-breaking 
parameters; it is needed for the invariance of the potential under
a RG transformation.}.
The tree-level potential $V_0$ is 
\begin{equation}
V_0(h_1,h_2)\ =\ 
{1\over2}(m^2_{H_1}+\mu^2)h^2_1+{1\over2}(m^2_{H_2}+\mu^2)h^2_2
+B_\mu h_1 h_2\ ,\label{tree}
\end{equation}
which simply follows from substituting Eq.~(\ref{h1h2}) into
the MSSM Higgs sector scalar potential (\ref{mssm}).

The one-loop potential is well known and the ${\cal O}(\alpha_s\alpha_t)$ 
part of the two-loop potential was computed in \cite{RenJie}; we list them 
here for completeness and for future reference. The complete
one-loop potential in Laudau gauge
is\footnote{We adopt the (modified) $\dr$-scheme of Ref.~\cite{DR}.}
\begin{eqnarray}
\olf~V_1(h_1, h_2)
 &=& \sum_{f} N_c^f \biggl[\sum_{i=1,2} H(m^2_{\tilde f_i}) - 2 
H(m_f^2)\biggr] + 3 H(m_W^2) + {3\over 2} H(m^2_Z) \nonumber\\ 
&+ &  {1\over 2}\sum_{n=1}^4 H(m^2_{H^0_n}) + \sum_{n=1}^2 H(m^2_{H^+_n})
- 2 \sum_{i=1}^2 H(m^2_{{\tilde\chi}^+_i}) 
- \sum_{i=1}^4 H(m^2_{{\tilde\chi}^0_i})\ ,
\label{1l} 
\end{eqnarray}
where $f$ sums over all the (s)quarks and (s)leptons,
$N_c^f$ is the color factor, 3 for (s)quarks and 1 for (s)leptons.
Following the leading approximation, 
we keep only the numerically important 
parts, {\it i.e.},
those from top (s)quarks. In Eq.~(\ref{1l}),
$\tilde{\chi}^+_i (i=1,2)$ and $\tilde{\chi}^0_i (i=1,2,3,4)$ 
represent charginos and neutralinos, and the function $H$ is 
\begin{equation}
H(m^2)={m^4\over 2}\biggl(\ln{m^2\over Q^2} - {3\over 2}\biggr)\ .
\end{equation}

The QCD contribution to the two-loop effective potential in the MSSM is
\begin{eqnarray}
&&(\olf)^2~V_{2s}(h_1,h_2)\ =\
8 g^2_3 \Biggl\{J(m_t^2,m_t^2)-2 m^2_t\ I(m_t^2,m_t^2,0) \nonumber\\
&&+{1\over 2}(c^4_t+s^4_t)\sum_{i=1}^2 J(m^2_{\tilde t_i},m^2_{\tilde
t_i})
+ 2 s^2_t c^2_t J(m^2_{\tilde t_1},m^2_{\tilde t_2})
+ \sum_{i=1}^2 m^2_{\tilde t_i}
I(m^2_{\tilde t_i},m^2_{\tilde t_i},0)\nonumber\\
&&+\sum_{i=1}^2 L(m^2_{\tilde t_i},m^2_{\tilde g},m^2_t)
-4 m_{\tilde g}\ m_t\ s_t c_t
\left[I(m^2_{\tilde t_1},m^2_{\tilde g},m^2_t)
-I(m^2_{\tilde t_2},m^2_{\tilde g},m^2_t)\right]\nonumber\\
&& + \left[m_t\rightarrow m_b,\; m_{\tilde t_i}\rightarrow m_{\tilde
b_i},\;
\theta_{\tilt}\rightarrow \theta_{\tilb}\right]
\Biggr\}\ ,
\label{2las}
\end{eqnarray}
where $\tilde g$ is the gluino, with tree-level mass given by the
$SU(3)$ gaugino soft mass $M_3$.  The last term, obtained by interchanging
variables, gives the contribution from sbottoms. Note that there is no  
mixed contribution involving stops and sbottoms, even if such mixed
couplings exists [from $SU(3)$ $D$-terms].
The
two-loop scalar functions $I$, $J$ and $L$ in Eq.~(\ref{2las}) 
are given in Appendix A, Eqs.~(\ref{I}), (\ref{J}) 
and (\ref{L}).\footnote{The procedure we have 
followed of subtracting all possible one-loop
subdivergences to define these functions is an alternative to the 
direct way used in Ref.~\cite{H2}. The direct way requires the
computation of some one-loop quantities to order ${\cal O}(\epsilon)$; 
perhaps we find the subtraction method simpler.
We have explicitly checked that, in the particular limit studied in
\cite{H2}, we exactly reproduce their unexpanded mass formula, Eq.~(11) of
\cite{H2}, which shows the equivalence of both methods. }

\begin{figure}[tbh]
\postscript{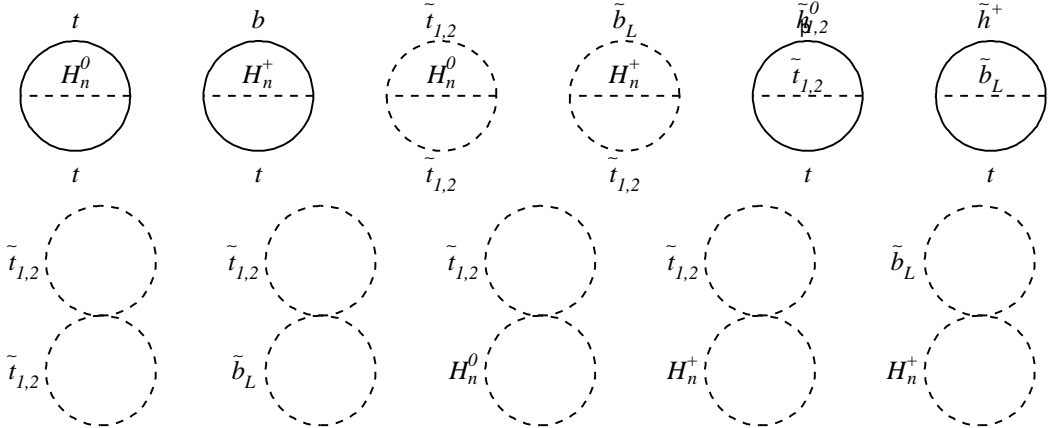}{1.0}
\caption{Feynman diagrams for the two-loop effective potential 
of order ${\cal O}(\alpha_t^2)$ in the MSSM. $H_n^0$
represent $H^0, h^0, G^0$ and $A^0$ for $n=1,2,3,4$. 
$H_n^+$ represent $G^+, H^+$ for $n=1,2$. The neutral and charged 
Higgsinos ${\widetilde h}_{1,2}^0~(\equiv {\tilde\chi}_{3,4}^0)$ 
and ${\widetilde h}^+~(\equiv {\tilde\chi}_{2}^+)$ 
have degenerate mass of $|\mu|$. }
\label{fig:Feyn}
\end{figure}

The top Yukawa contribution to the two-loop potential is a new
result of this paper. The relevant Feynman diagrams are shown in 
Fig.~\ref{fig:Feyn}. To simplify the final result, we neglect left-right
mixings in the bottom-squark sector and the gaugino-Higgsino mixings in the
neutralino-chargino sector (under this assumption, the Higgsino
masses are simply $|\mu|$); these simplifications are valid in the 
leading approximation. Using the Feynman rules given in Appendix B,
we find (the last diagram of Fig.~\ref{fig:Feyn} is of order $h_t^2$
but does not contribute to $m_{h^0}$)
the top and bottom Yukawa contribution to the two-loop potential\footnote{
In this revised version, we have also included the bottom Yukawa
contributions for completeness. All analyses in the main text use 
only the top Yukawa contributions as in the previous version.}:
\begin{eqnarray}
&&(\olf)^2~(V_{2t}(h_1,h_2)+V_{2b}(h_1,h_2))\ =\ \nn\\
&&\left[3 h_t^2\Biggl\{\sum_{n=1}^4\frac{D_n}{2}\biggl[L(m^2_{H^0_n},
m^2_t,m^2_t)
\pm 2 m^2_t I(m^2_{H^0_n}, m^2_t,m^2_t) 
+ \sum_{i=1}^2 J(m^2_{\tilt_i},m^2_{H^0_n})\biggr]\right.\nn\\
&&+\sum_{n=1}^2 D_{n+2}\biggl[s^2_t J(m^2_{\tilt_1},m^2_{H^+_n})
+c^2_t J(m^2_{\tilt_2}, m^2_{H^+_n})+c^2_bJ(m^2_{\tilb_1}, m^2_{H^+_n})+
s^2_bJ(m^2_{\tilb_2}, m^2_{H^+_n})
\biggr]\nn\\
&&+s^2_t\biggl[c_b^2 J(m^2_{\tilt_1},m^2_{\tilb_1})+
s_b^2 J(m^2_{\tilt_1},m^2_{\tilb_2})\biggr] 
+ c^2_t \biggl[c_b^2J(m^2_{\tilt_2},m^2_{\tilb_1})+
s^2_b J(m^2_{\tilt_2},m^2_{\tilb_2})\biggr]\nn\\
&&+s^2_{2t}\sum_{i=1}^2 J(m^2_{\tilt_i},m^2_{\tilt_i})
+c_{4t}J(m^2_{\tilt_1},m^2_{\tilt_2})+ L(m^2_{\tilt_1},\mu^2,m_t^2)
+ L(m^2_{\tilt_2},\mu^2,m_t^2) \biggr\}\nn\\
&&+3\biggl\{(h_t^2s_t^2+h_b^2c_t^2)L(m^2_{\tilt_1},\mu^2,m_b^2)
+(h_t^2c_t^2+h_b^2s_t^2)L(m^2_{\tilt_2},\mu^2,m_b^2)\nn\\
&&-2\mu m_b h_t h_b 
s_{2t}\biggl[I(m^2_{\tilt_1},\mu^2,m_b^2)-I(m^2_{\tilt_2},\mu^2,m_b^2)
\biggr]\biggr\}\nn \\
&&\left.+\left\{
h_t\rightarrow h_b,\;
m_t\lra m_b,\;
m_{\tilde t_i}\lra m_{\tilde b_i},\;
\theta_{\tilt}\lra \theta_{\tilb},\;
D_k\rightarrow D'_k\right\}\right]\nn\\
&&+3 \biggl\{
(h_t^2s_\beta^2+h_b^2c_\beta^2)L(m^2_{G^+}, m^2_t,m^2_b)
+(h_t^2c_\beta^2+h_b^2s_\beta^2)L(m^2_{H^+}, m^2_t,m^2_b)\nn\\
&&+2m_tm_bh_th_bs_{2\beta}\biggl[I(m^2_{G^+}, m^2_t,m^2_b)
-I(m^2_{H^+}, m^2_t,m^2_b)\biggr]\biggr\}\nn\\
&&- \frac{3}{2}\sum_{i,j=1}^2\sum_{n=1}^4 \lambda^2_{H^0_n\tilf_i\tilf_j}
I(m^2_{H^0_n},m^2_{\tilf_i},m^2_{\tilf_j})
-3\sum_{i,j=1}^2\sum_{n=1}^2 \lambda^2_{H^+_n\tilf_i\tilf_j}
I(m^2_{H^+_n},m^2_{\tilf_i},m^2_{\tilf_j})\ ,
\label{2lat}
\end{eqnarray}
where, in the first line of Eq.~(\ref{2lat}), 
positive and negative signs apply to ${\cal CP}$-even ($H^0,~h^0$) 
and odd ($A^0,~G^0$) Higgs/Goldstone bosons respectively, and in the last 
line $\tilf_i=\{\tilt_i,\tilb_i\}$.
We also have, for $n=1,2,3,4$, $D_n=\{\sa^2,\ca^2,\sbe^2,\cbe^2\}$  
and $D'_n=\{\ca^2,\sa^2,\cbe^2,\sbe^2\}$. The ordering of the
Higgs/Goldstone bosons is $H_n^0=H^0, h^0, G^0$ and $A^0$ for $n=1,2,3,4$
and $H_n^+=G^+, H^+$ for $n=1,2$.

Two tests can be applied to check the correctness of the effective
potential $V(h_1,h_2)$. First, the potential should vanish 
in the supersymmetric limit ({\it i.e.}, when all soft-breaking parameters 
are taken to be zero),
and second, the potential $V(h_1,h_2)$ should be
invariant under changes of the renormalization
scale, up to the order of our perturbative calculation.
The vanishing of the potential in the supersymmetric limit is 
proved by simple algebra. In the following we show the
invariance of the two-loop potential under a RG transformation.

Using the derivatives of $I$, $J$ and $L$ functions
with respect to the renormalization scale $Q$ 
\begin{eqnarray}
&&{\partial I(m_1^2,m_2^2,m_3^2)\over\partial\ln Q^2}
\ =\ -\sum_{i=1}^3\biggl[A_0(m_i^2)+m_i^2\biggr]\ ,\\
&&{\partial J(m_1^2,m_2^2)\over\partial\ln Q^2}
\ =\ m_1^2 A_0(m_2^2) + m_2^2 A_0(m_1^2)\ ,\\
&&{\partial L(m_1^2,m_2^2,m_3^2)\over\partial\ln Q^2}
\ =\  (m_1^2-2m^2_2-2 m^2_3) A_0(m_1^2)
-m^2_2 A_0(m_2^2) - m^2_3 A_0(m_3^2)\nn\\
&&\qquad\qquad\qquad\qquad~~ +m_1^4 - (m_2^2+m_3^2)^2\ , 
\end{eqnarray}
and the one-loop MSSM RGEs for top-(s)quark and Higgs boson masses,
Eqs.~(\ref{mt}-\ref{H+}), we find that the RG variation 
\begin{eqnarray}
-{\partial V_{2}\over\partial\ln Q^2}
-{\cal D}^{(1)}V_{1} &=&
{8g_3^2h_t^2h_2^2\over(16\pi^2)^2}
\biggl(\MQ^2+\MU^2+2M_3^2+\xt^2
-2M_3\xt\biggr)\nn\\
&-&{9h_t^4 h_2^2\over(\olf)^2}\biggl\{\MQ^2+\MU^2+{1\over2}\biggl[
m^2_{H_2}+(A_t+\xt)^2\biggr]\biggr\}
\label{eq:rgv}
\end{eqnarray}
modulo terms independent of the Higgs field $h_2$. Here
${\cal D}^{(1)}V_{1}$ represents 
the one-loop RG variation of the one-loop potential Eq.~(\ref{1l}). 
This result agrees exactly with 
the two-loop RG variation of the tree-level potential ${\cal D}^{(2)}V_{0}$
[{\it cf.} Eqs.~(\ref{tree}) and  (\ref{mH2}-\ref{bmu})], so that
\be
\frac{d }{d\ln Q^2}(V_0+V_1+V_2)\equiv
{\cal D}^{(2)}V_0+
{\cal D}^{(1)}V_1+
\frac{\partial V_2}{\partial \ln Q^2}=0\ .
\ee
Note that this is a nontrivial check 
that all $\ln Q^2$ terms cancel with each other between Eq.~(\ref{eq:rgv})
and ${\cal D}^{(2)}V_{0}$; this cancellation 
guarantees the correct leading and 
next-to-leading order logarithmic behavior of the effective potential.

To derive the analytical expression of $\Delta m_{h^0}^2$ in Sec. 2,
we need to expand the two-loop potential $V_2$ in powers of
$m_t/M_S$ and $m_t X_t/M_S^2$; besides many straightforward
expansions, we have used (\ref{xiapp1})-(\ref{xiapp2}) 
for the $t-\tilde{q}-\tilde{h}$  and
$\tilde{t}-\tilde{q}-h$ diagrams of (\ref{2lat}), with
$\tilde{q}=\tilde{t}$ or $\tilde{b}$.

\end{document}